\documentclass[12pt]{article}

\usepackage[margin=1in]{geometry}
\usepackage{lmodern}

\usepackage{amsmath,amssymb,amsfonts,dsfont}
\DeclareMathOperator*{\argmin}{arg\,min}
\DeclareMathOperator*{\argmax}{arg\,max}

\usepackage{graphicx}
\usepackage{float}
\usepackage{listings}
\usepackage{booktabs}
\usepackage{array}
\usepackage{multirow}
\usepackage{bigstrut}
\usepackage{longtable}
\usepackage{threeparttable}
\usepackage{caption}
\usepackage{subcaption} 
\usepackage{pdflscape}
\usepackage{tikz}
\usetikzlibrary{positioning, arrows.meta}

\usepackage[dvipsnames]{xcolor}
\definecolor{codebg}{HTML}{FBF7EF}
\definecolor{coderule}{HTML}{E3D7BE}
\definecolor{codekeyword}{HTML}{1D5C63}
\definecolor{codeemph}{HTML}{0B4F6C}
\definecolor{codestring}{HTML}{B04A2A}
\definecolor{codecomment}{HTML}{6C7A5B}
\lstdefinestyle{policytargetr}{
  language=R,
  basicstyle=\ttfamily\small,
  backgroundcolor=\color{codebg},
  frame=single,
  rulecolor=\color{coderule},
  framesep=8pt,
  xleftmargin=0.02\linewidth,
  xrightmargin=0.02\linewidth,
  keywordstyle=\color{codekeyword}\bfseries,
  emph={library,analyze_rectangular_policy,plot_policy_values,save_policy_outputs},
  emphstyle=\color{codeemph}\bfseries,
  stringstyle=\color{codestring},
  commentstyle=\color{codecomment}\itshape,
  showstringspaces=false,
  breaklines=true,
  keepspaces=true,
  columns=fullflexible,
  aboveskip=0.8em,
  belowskip=0.4em
}
\usepackage{hyperref}
\hypersetup{
  colorlinks=true,
  citecolor=blue,
  urlcolor=blue,
  linkcolor=blue
}

\usepackage[authoryear,longnamesfirst,round]{natbib}
\usepackage{amsthm}

\newtheorem{assumption}{Assumption}
\newtheorem{definition}{Definition}
\newtheorem{proposition}{Proposition}
\newtheorem{remark}{Remark}
\newtheorem{example}{Example}

\usepackage{algorithm}
\usepackage{algpseudocode}

\usepackage{tikz-cd}
\usepackage{afterpage}
\usepackage{comment}
\usepackage{footmisc}
\usepackage{chngcntr}

\title{Producing Policy Recommendations: from Statistical Decision Theory to Empirical Practice}
\author{Giacomo Opocher\thanks{\scriptsize University of Bologna, giacomo.opocher2@unibo.it. \textbf{This version is a work in progress; please feel free to contact me if you find a mistake, think I missed an important contribution, or just want to send a comment.} This paper has greatly benefited from the guidance of Silvia Sarpietro and insightful discussions with Marc Clos and Juan C. Yamin. All mistakes are my own.}}
\date{\today}
\begin{document}
\maketitle
\vspace{-1.7cm}
\begin{abstract}
    \textbf{Abstract}. 
    Applied research in economics is intrinsically motivated by broad normative objectives. 
    However, it is not obvious how a researcher should direct their efforts to produce evidence toward such objectives.
    This paper reviews recent theoretical developments on research design for policy choice and provides new tools applied researchers can use to guide their design choices and communicate their policy recommendations.
    First, I focus on theoretical contributions in econometrics and provide a general framework that nests all the contexts and results reviewed using a coherent notation and narrative.
    %I discuss in detail the theoretical fabric most results rely on and the types of guarantees one can achieve. 
    Then, I present two diagrams applied researchers can use to navigate the theoretical literature starting from concrete scenarios to make thoughtful design choices.
    Finally, I introduce a new \texttt{R} package that produces one table and two figures applied researchers can plug in their `policy implications' section to provide evidence on the performance of different policy recommendations coming out of their study. 
    The use of such tools is illustrated with an example in development economics. 
    \vspace{2cm}
\end{abstract}

\section{Introduction}\label{sec:intro}
Around the world, public institutions draw on economic research to guide real-world policy choices with empirical evidence.

In the United States, the \hyperlink{http://whitehouse.gov/cea/}{Council of Economic Advisers} is `charged with offering the President objective economic advice on the formulation of [...] economic policy [...] based [on the] analysis [of] economic research and empirical evidence'.
In the European Union, `[the main objective of the] economic research activities of the \hyperlink{https://economy-finance.ec.europa.eu/economic-research-and-databases/economic-research_en?prefLang=de}{European Commission}, [...] is to support policy making [...] by developing research tools and analysing data'.
In Japan, the \hyperlink{https://www.esri.cao.go.jp/en/esri/about/menu-e.html?utm_source=chatgpt.com}{Economic and Social Research Institute} mainly engages in [...] conducting empirical research related to economic and social activities [...] [to promote] evidence-based policy making.

Research institutions, academic associations, and researchers acknowledge this role.

The \hyperlink{https://www.nber.org/research}{National Bureau of Economic Research} identifies as its main objective `conducting and disseminating independent, cutting-edge, non-partisan research that advances economic knowledge and informs policy makers and the business community'.
The \hyperlink{https://cepr.org/about}{Center for Economic Policy Research} claims its main objective is `to enhance the quality of economic policy-making within Europe and beyond, by fostering high quality, policy-relevant economic research'.
The \hyperlink{https://www.aeaweb.org/about-aea}{American Economic Association} defines its members as `committed to the advancement of economics and its enduring contributions to society'.
Indeed, around 70\% of the empirical papers published in an AEA journal in the last 10 years provide at least one policy recommendation.
The total number of recommendations is 517 across 471 empirical papers, 321 of which have at least one recommendation.
As a result, the average empirical paper carries between one and two policy recommendations.\footnote{Author's calculations based on the full text of the universe of papers published in AER, AER:Insights, AEJ:Applied, AEJ:EconPolicy, and AER:P\&P from 2015 to 2025. See Appendix \ref{app:intro_counts_methodology} for details.}

Therefore, at least in intention, academic papers in economics are motivated (and often funded) by broader normative objectives. 

By the nature of many of the questions we ask across fields of economics, such objectives can be framed as the goal of understanding how to \textit{intervene} on a given status quo to increase welfare.
Examples include understanding whether introducing a minimum wage reduces or increases unemployment \citep{card_krueger_2000}; 
understanding whether providing free health insurance improves access to health care services \citep{medic_aid_2016};
understanding whether exposing poor families to better neighborhoods improves children’s long-run outcomes \citep{moving_to_opp}.

The success of such ambitious objectives rests on the premise that economists' \textit{research design} choices are aligned with them.

Providing answers to policymakers, that is, producing policy recommendations on a given set of interventions of interest, requires making design choices along at least three dimensions.
(i) Planning data collections (e.g. what individual characteristics are relevant, how many individuals should be sampled, what measures should be considered, etc.), 
(ii) choosing an estimator to provide sufficient statistics to compute the \textit{value} of each policy, 
(iii) defining a decision rule, that is, a function that ranks different policies according to their estimated value.
It is not obvious a priori how to formalize the problem of optimally learning policy recommendations, what criterion a research design should maximize to be considered optimal or aligned with the broader normative objective, and what criterion the mainstream practice is implicitly maximizing.

The problem of making such choices becomes ever more challenging if they need to be made \textit{ex-ante}, meaning when only minimal or no information about (i) the probability laws of the population of interest and (ii) the causal law of a given intervention is available.
Ideally, researchers would like to commit to a choice of the three elements listed above such that they are certain that, even in the \textit{worst case} admitted by those minimal assumptions, such choices will not be \textit{too far} from being optimal. 

For example, for a given estimator and choice rule, would it be preferable to collect data from a lab or a field experiment? 
At what level should the intervention be stratified, and how many units should be collected for each stratum?
What is the minimum experimental sample size that would guarantee that our final recommendation is \textit{at most as far} from optimal?
Similar questions can be framed while keeping the collection plan fixed and searching for an optimal estimator or choice rule.

A recent literature initiated by \cite{manski_statistical_2004} has examined these problems through the lens of statistical decision theory \citep{wald1950statistical,Savage01031951}.
As the literature has arguably achieved theoretical maturity, this paper contributes to it by reviewing the main theoretical contributions under a common general framework, identifying the directions in which recent research is moving and questions on the frontier.\footnote{An important remark is that in this review I focus on the econometrics literature on the topic and exclude the computer science and biostatistics literatures from the scope.
Nevertheless, I think it would be of great value to reconcile these different branches.}
Moreover, I provide guidelines for practitioners on how to leverage such results to make ex-ante choices about research designs when the ultimate goal is to provide policy recommendations, and operational tools to communicate the performance of the resulting recommendation once the data are realized.

\paragraph{Structure of the Review.} 
The review is organized in two main parts. 
The first is theoretical. 
Section \ref{sec:general_framework} introduces a common decision-theoretic framework and formalizes the research design problem for policy choice. 
Section \ref{sec:policy_choice} then reviews the main theoretical results, first for unconstrained policy spaces (Section \ref{sec:unconstrained}) and then for constrained policy spaces (Section \ref{sec:constrained}). 
Section \ref{sec:interesting_venues} concludes the theory part by discussing frontier directions, with particular attention to evidence aggregation and implementation. 

The second part is applied. 
Section \ref{sec:how_to_produce} provides (i) a map (Figures \ref{fig:unconstrained_tree} and \ref{fig:constrained_tree}) for applied researchers to navigate the theory starting from concrete concerns when deciding among research designs (Section \ref{sec:before_collecting}), and (ii) a workflow to solve the ex-post problem of communicating the performance of different recommendations once the data are realized (Section \ref{sec:after_collecting}).
For the latter purpose, I introduce (the beta version of) a new \texttt{R} package named \texttt{policytargetr} that practitioners can use to evaluate the performance of their recommendations before recommending them.
Both objectives are illustrated in the setting of \citet{hussam_2022}.

\section{General Framework}\label{sec:general_framework}
Given a state $Q \in \mathcal{Q}$, a data-collection design $\omega \in \Omega$, and design parameters $\theta \in \Theta_\omega$, let $Q_{\omega,\theta}$ denote the probability law over the sample space $\mathcal{S}$ induced by implementing $(\omega,\theta)$ under state $Q$. The observed sample therefore satisfies
\begin{equation}
    S \sim Q_{\omega,\theta}.
\end{equation}

An example is the collection of a panel dataset ($\omega$) that takes the cross-sectional sample size (e.g. \# individuals, $n$) and the longitudinal dimension (e.g. \# time periods, $t$) as inputs, $\theta = (n,t)$. 
Another example is the collection of data from a randomized controlled trial ($\omega$) that takes the share of units that receive the intervention ($p$) and the sample size as inputs ($n$): $\theta = (n, p)$.
Data collections are costly.
Costs vary with the choice of the collection design ($\omega$), the design's parameters ($\theta$) and the probability law $Q$.
Define a cost function $c(\omega,\theta,Q)$:
\begin{equation}\label{eq:cost_function}
    c: \Omega \times \Theta_\omega \times \mathcal{Q} \to \mathbb{R}^+.
\end{equation}
As an example, if $\omega$ is a survey experiment, and the intervention is costly to provide, different choices of $\theta = (n,p)$ map into different costs.
Next, given a set of policies $\mathcal{D}$ with typical element $d$, define a set of estimators $m(S,d)$ as:
\begin{equation}\label{eq:estimator}
    \mathcal{M} := \{m:\mathcal{S} \times \mathcal{D} \to \mathcal{Z}\},
\end{equation}
where $\mathcal{Z}$ denotes a general evidence set. 
Depending on the application, an element of $\mathcal{Z}$ may include point estimates, confidence sets, standard errors, combinations of these objects, or other statistics relevant to policy choice.
As an example, consider an experimental data collection (defining $\omega$) with $n$ units and propensity score $p$ (defining $\theta$).
Let $\mathcal{M}$ denote a set of estimators for the expected outcome in case of intervention, or at the status quo. 
One of such estimators is the empirical mean.
Let $\mathcal{D} = \{0,1\}$ denote the choice of whether to introduce an intervention or not.
Note that in this simple case the policy set coincides with the interventions set, but in general it needs not to be the case.
In this example $\mathcal{Z}= \mathbb{R}$ contains the estimated expected outcome.

Given the estimator, denote a set of choice rules $\delta(m(S,\cdot))$, where $m(S,\cdot):=\{m(S,d): d \in \mathcal{D}\}$, as:
\begin{equation}\label{eq:decision_rule}
    \Delta := \{\delta: \mathcal{Z}^{\mathcal{D}} \to \mathcal{D}\}.
\end{equation} 
Following up on the previous example, one simple choice for this rule is the following:
\begin{equation}
\delta(m(S,\cdot))= \mathbf{1}\{m(S,1)> m(S,0)\}
\end{equation}
I report in Table \ref{app:tab:notation_applied} other examples for each margin of choice.

In the following definition I state formally the general research design problem for policy choice, at a fixed state $Q$.

\begin{definition}[Policy Choice Decision Problem]\label{def:policy_choice_problem}
    A researcher interested in policy choice solves: 
    \begin{align}\label{eq:policy_choice_problem}
        & (\hat{\omega}, \hat{\theta}, \hat{m}, \hat{\delta}) = \arg \min_{\omega \in \Omega, \theta \in \Theta_\omega, m \in \mathcal{M}, \delta \in \Delta}  r\{Q,\omega,\theta,m,\delta\} \\
        & \text{\emph{subject to}}: c(\omega,\theta,Q)  \leq B_0,
    \end{align}
    where $r \in \mathcal{R}$ denotes a policy choice loss function $r : \mathcal{Q} \times \Omega \times \Theta_\omega \times \mathcal{M} \times \Delta \to \mathbb{R}^+$, and $B_0 \in \mathbb{R}^+$.
    The policy recommendation is then defined as:
    \begin{equation}
        \hat{d} = \hat{\delta}\left[\hat{m}\left(S, \cdot \right)\right], \qquad S \sim Q_{\hat{\omega},\hat{\theta}}
    \end{equation}
\end{definition}

\begin{remark}
    The researcher is deciding ex-ante the research design that will produce a policy recommendation given the data, not the recommendation itself. 
    The latter is just a function of the data and the design.
    Therefore, we seek theoretical guarantees on the design choice, not on the specific recommendation. 
\end{remark}

\paragraph{Dealing with Uncertainty of the State.} Depending on the context, the researcher may have different information about $\mathcal{Q}$, commonly called the state space \citep[see e.g.][]{manski_2021}. 
Various approaches have been proposed to leverage the knowledge available about $\mathcal{Q}$ to solve \textit{optimally} the researcher's problem.

For example, a researcher may know a probability law $\pi(Q)$ that assigns a probability to each $Q \in \mathcal{Q}$ and then solve:
\begin{align}\label{eq:bayes}
    & (\hat{\omega}, \hat{\theta}, \hat{m}, \hat{\delta}) = \arg \min_{\omega \in \Omega, \theta \in \Theta_\omega, m \in \mathcal{M}, \delta \in \Delta}  \mathbb{E}_{\pi}\Bigg[ r\{Q,\omega,\theta,m,\delta\} \Bigg] \\
    & \text{\emph{subject to}}: \mathbb{E}_\pi[c(\omega,\theta,Q)]  \leq B_0.
\end{align}
This is often referred to as the Bayesian approach \citep[see, e.g.][]{rubin_78,DEHEJIA2005141}.

A different approach is agnostic about $\pi(Q)$ and focuses on delivering uniform guarantees over the state space by minimizing the worst-case. 
Formally, the researcher solves:
\begin{align}\label{eq:minimax}
    & (\hat{\omega}, \hat{\theta}, \hat{m}, \hat{\delta}) = \arg \min_{\omega \in \Omega, \theta \in \Theta_\omega, m \in \mathcal{M}, \delta \in \Delta}  \max_{Q \in \mathcal{Q}} r\{Q,\omega,\theta,m,\delta\} \\
    & \text{\emph{subject to}}: \max_{Q \in \mathcal{Q}}c(\omega,\theta,Q)  \leq B_0.
\end{align}
This is known as the minimax approach, introduced as a criterion for policy choice problems by \cite{manski_statistical_2004} and first theorized in generality by \cite{wald1950statistical} and \cite{Savage01031951}.

Another approach derives $Q$-dependent guarantees by setting parametric constraints on $\mathcal{Q}$ via a parameter space $\beta \in \mathcal{B}$ such that $\{Q_\beta : \beta \in \mathcal{B}\} \subseteq \mathcal{Q}$ and solves:
\begin{align}
    & (\hat{\omega}, \hat{\theta}, \hat{m}, \hat{\delta}) = \arg \min_{\omega \in \Omega, \theta \in \Theta_\omega, m \in \mathcal{M}, \delta \in \Delta}  r\{Q_\beta,\omega,\theta, m, \delta\} \\
    & \text{\emph{subject to}}: c(\omega,\theta,Q_\beta) \leq B_0.
\end{align}

In this review, I focus on the minimax approach, as it is the approach most widely used by the theoretical literature on policy choice following \cite{manski_statistical_2004}.

\begin{remark}\label{rmk:two_assumptions}
    The mainstream application of the minimax principle to policy choice problems imposes two implicit assumptions.
    First, the choice of $(\omega, \theta,m,\delta)$ happens before any information about $Q$ other than the restrictions that define a feasible state space $\mathcal{Q}$ is available. 
    Second, the estimating distribution (i.e. the one from which the learning sample is drawn) and the target distribution (i.e. the one in which the policy will be applied) coincide.
    Some relaxations of these two assumptions \citep[e.g.][]{christensen_payoffs_2025} have been considered and are discussed in this review.
\end{remark}

\begin{remark}\label{rmk:general_specific}
    Although in the general framework presented here the choice of $(\omega, \theta,m,\delta)$ is endogenous, theoretical papers usually specify one pair $(m,\delta)$ (e.g. conditional empirical success rules in \cite{manski_statistical_2004} or empirical welfare maximization in \cite{kitagawa_who_2018}) and derive its theoretical properties, conditional on $(\Omega,\Theta_\omega)$ satisfying a set of assumptions (e.g. unconfoundedness of the intervention assignment and strict overlap in \cite{kitagawa_who_2018}), rather than providing a criterion for finding the joint optimum.
    Other approaches define general properties that $m$ should satisfy (e.g. asymptotic normality as in \citet{hirano_asymptotics_2009}) and search for an optimal decision rule $\delta$.
\end{remark}

\paragraph{Choice of the Loss Function.} The choice of $r$ is kept exogenous, or rather decided by the econometrician to provide theoretical results.
The mainstream approach in the literature is to consider expected regret. 
Defining regret requires two additional definitions. First, a welfare functional
\begin{equation}
    W : \mathcal{D} \to \mathbb{R},
\end{equation}
which is typically evaluated at the expected value $W_Q(d) = \mathbb{E}_Q[W(d)]$. 
Next, we need to define an oracle that knows $W_Q(d)$, and therefore can directly solve:
\begin{equation}
    d^* = \argmax_{d \in \mathcal{D}} W_Q(d).
\end{equation}
Then, we can define regret as:
\begin{equation}\label{eq:regret}
    R_Q(\hat d) := \mathbb{E}_{Q_{\omega,\theta}}[W_Q(d^*) - W_Q(\hat{d})],
\end{equation}
where $\hat{d}$ implicitly carries forward the choice of $(\omega,\theta, m, \delta)$, since they are necessary to compute it.
For expositional convenience, I use $\hat{d}$ as a shortcut for $(\hat{\omega}, \hat{\theta}, \hat{m}, \hat{\delta})$. 
However, it must be clear that what I really mean by $\hat{d}$ is the whole research design, and not the actual element $d \in \mathcal{D}$ (see Remark \ref{rmk:not_d} for further details).
Regret measures the average welfare loss from recommending the estimated policy $\hat d$ instead of the oracle policy $d^*$, over draws of the estimating sample $S^n$.
The outer expectation is taken with respect to $Q_{\omega,\theta}$, the sampling law induced jointly by the state $Q$ and the collection design $(\omega,\theta)$.

\begin{remark}
    The design decision problem from the minimax regret perspective can be interpreted as the search for a research design that, even in the worst admissible state of nature, would provide a recommendation that is not too far from the best one could possibly give.
    Importantly, this is not a guarantee about the realized recommendation itself, but about the way of producing it.
\end{remark}

\begin{remark}\label{rmk:not_d}
    The outer expectation in Eq. \ref{eq:regret} defines the loss as the average welfare loss over infinitely many draws of the estimating sample, given the probability law $Q$ and the sampling design $(\omega,\theta)$.
    As a result, this guarantee only informs the researcher about the loss that the average sample would produce if we were to proceed with that research design. 
    This definition reflects the ex-ante perspective described above: we make our choices based on what we expect to happen if we run one design or another.
    Therefore, this approach does not inform the policymaker about the performance of the recommendation that the specific sample drawn will produce. 
\end{remark}

\begin{remark}
    It is not immediate to provide a concrete meaning to regret. 
    One interpretation is provided by the following thought experiment. 
    Assume the policymaker has a finite budget and needs to compensate for any wrong decision (meaning, giving back to a misassigned unit the absolute value of the intervention's effect). 
    Then, worst-case regret is the maximum compensation cost a policymaker may have to pay.
    Although this compensation cost may not be a real quantity (e.g. because it cannot be expressed in monetary units, or because there is no way to identify a mistake), at least it provides a uniform measure to rank different research designs whose value would otherwise be hard to rank.
\end{remark}

\paragraph{Types of Theoretical Guarantees.} 
Given a choice of $(\omega,\theta, m, \delta)$, computing a (non-trivial) closed-form expression for $\sup_{Q \in \mathcal{Q}} R_Q(\hat{d})$ under minimal assumptions on $\mathcal{Q}$ is often not possible.\footnote{The formal discussion of these \textit{minimal} assumptions is deferred. They often entail a uniform bound on the outcome, conditional independence of the intervention assignment, and strict overlap \citep[see, e.g.][]{kitagawa_who_2018}.}
As a result, it is mainstream theoretical practice to derive upper and lower bounds for worst-case regret:
\begin{equation}\label{eq:lower_upper}
\underline{\mu}(\hat{d}) \leq \sup_{Q \in \mathcal{Q}} R_Q(\hat d) \leq \bar{\mu}(\hat{d}).
\end{equation}
Deriving upper bounds can be easy. The theoretical crux comes with proving that the derived upper bound is sharp, meaning that it is attained with equality for an admissible DGP.
For a given research design $\hat d$, this can be shown by finding a specific $Q' \in \mathcal{Q}$ such that
\begin{equation}\label{eq:sharp_bound}
R_{Q'}(\hat d) = \bar{\mu}(\hat d).
\end{equation}
Sharpness of the upper bound identifies the exact worst-case regret of that research design. To additionally establish minimax optimality, one needs to show that no alternative research design can attain a lower worst-case regret. In particular, letting $\hat d^*$ denote a minimizer of $\bar{\mu}(\hat d)$, it is sufficient to establish the minimax lower bound
\begin{equation} \label{eq:minimax_opt}
\min_{m \in \mathcal{M}, \delta \in \Delta}
\sup_{Q \in \mathcal{Q}} R_Q(\hat d)
\geq
\bar{\mu}(\hat d^*).
\end{equation}
Together with the upper bound for $\hat d^*$, this implies that $\hat d^*$ is minimax-optimal. Such lower bounds can often be established by constructing specific DGPs, or families of DGPs, for which no alternative algorithm $(m',\delta')$, given $(\omega,\theta)$, can achieve lower worst-case regret.

Unfortunately, in many cases, it is not possible to establish minimax optimality exactly. This could be because the upper bound is not sharp, or because a matching minimax lower bound cannot be derived.
The second-best result, then, is to prove that the rate of the upper bound is minimax-sharp. In particular, for some sequence $a_n$, suppose that
\begin{equation}\label{eq:minimax_rate_opt}
\sup_{Q \in \mathcal{Q}} R_Q(\hat d^*) \leq C a_n 
\qquad \text{and} \qquad
\min_{m \in \mathcal{M}, \delta \in \Delta}
\sup_{Q \in \mathcal{Q}} R_Q(\hat d) \geq c a_n.
\end{equation}
Then no alternative research design can improve on the rate $a_n$, and $\hat d^*$ is minimax rate-optimal.
One way to achieve this result is to prove that the upper and minimax lower bounds coincide up to constants.

Another result that can be stated when both upper and lower bounds are available is minimax dominance.
In particular, if the lower bound for a choice $\hat{d}'$ is higher than the upper bound for a choice $\hat{d}''$, then it must be that $\hat{d}''$ is minimax-dominant compared to $\hat{d}'$.
Formally,
\begin{equation}\label{eq:minimax_dom}
\text{If} \quad \underline{\mu}(\hat{d}') > \bar{\mu}(\hat{d}'')
\quad \Rightarrow \quad
\arg \min_{{\hat{d}'', \hat{d}'}}
\max_{Q \in \mathcal{Q}} R_Q(\hat{d}) = \hat{d}''.
\end{equation}
This is a sufficient condition for $\hat d''$ to be preferred to $\hat d'$ according to the minimax criterion.

\begin{remark}
    When sharpness or rate-sharpness is proven by examples, it is often the case that $Q'$ is a degenerate DGP.
    Since such DGPs are sometimes implausible in many empirical settings, this is often mentioned as a weakness of the minimax approach \citep[a similar argument is reported in][]{stoye_covariates_2012}.
    This is the cost of being as general as possible with the restrictions on $\mathcal{Q}$, combined with adopting an ex-ante perspective.
    If the $Q'$ that proves sharpness is not possible given a first draw of $n$ units, the upper bound may no longer be sharp after the data are realized, and the minimax optimality of $\hat{d}$ may be undermined.
    One alternative approach could be to update the worst case conditional on the family of distributions that remain possible after the data are realized, blending minimax and Bayesian thinking.
\end{remark}

\section{Policy Choice Decision Problems}\label{sec:policy_choice}
In what follows, I first specialize the general framework proposed in Section \ref{sec:general_framework} to the potential outcome framework, as this is the causal language considered in the literature, and then review the main theoretical contributions.

Let $\mathcal{T}$ denote a finite set of interventions and let $\mathcal{X}$ denote the $s$-dimensional space of observed covariates. 
Then, let $d : \mathcal{X} \to \mathcal{T}$ denote the assignment rule. 
Denote the set of assignment rules considered by $\mathcal{D}$.
A state $Q \in \mathcal{Q}$ specifies the joint distribution of $(T_i, X_i,\{Y_i(t)\}_{t \in \mathcal{T}})$.
For a given design $\theta$, the data collection process $(\omega,\theta)$ generates a sample
\begin{equation}
    S^n := \{(X_i,T_i,Y_i)\}_{i=1}^{n} \sim Q_{\omega,\theta},
\end{equation}
where $T_i$ denotes intervention assignment and $Y_i = Y_i(T_i)$ the observed outcome. 

\subsection{Unconstrained Policy Spaces}\label{sec:unconstrained}
\cite{manski_statistical_2004} was the first to apply statistical decision theory \citep{Savage01031951} to problems of policy choice.
He studies the problem of targeting interventions across subgroups of the covariate space when experimental data from the target population can be collected. 
He provides worst-case regret guarantees for \textit{conditional empirical success} (CES) rules, namely rules that target interventions to those subgroups that attain a higher empirical mean outcome when receiving the intervention.
The definition of subgroups plays a central role in the paper. 
These are defined through a coarsening $v:\mathcal{X} \to \mathcal{V}$, which aggregates covariate values into subgroups. 
A first conceptual point in \cite{manski_statistical_2004} is that, although an oracle policymaker would always prefer the finest coarsening to get as close as possible to individual variation in intervention effects, a statistical policymaker should choose the coarsening carefully to balance how well the assignment rule fits individual variation and how hard it is to learn from data.
This point highlights that standard \textit{bias-variance} tradeoffs in statistical learning can be studied analytically via worst-case regret in policy learning.
A second conceptual point is that data collection interacts with this tradeoff. 
Referring to the general formulation in Definition \ref{def:policy_choice_problem}, \citet{manski_statistical_2004} solves the joint problem of deciding subgroup definition (whether through the coarsening $v$ or the entire covariate space) and sample sizes ($\theta$ in Def. \ref{def:policy_choice_problem}'s notation) to attain the smallest worst-case regret possible. 
The key result is that there exist finite sample sizes such that it is minimax dominant (in the sense of Eq. \ref{eq:minimax_dom}) to consider the finest coarsening rule.

\cite{manski_statistical_2004} constrains the state space $\mathcal{Q}$ with the following assumptions.

\begin{assumption}[\cite{manski_statistical_2004}'s state space] \label{ass:manski_2004} $ $
    \begin{enumerate} 
        \item[\textbf{BO}.] The outcome variable satisfies $|Y_i(t)| \leq M/2$ for all $t \in \mathcal{T}$.
        \item[\textbf{CS}.] The covariate space is finite, $|\mathcal{X}| < \infty$; it has full support, $\mathbb{P}_Q(X_i = x)>0$ for all $x \in \mathcal{X}$; and these probabilities are known.
        \item[\textbf{AM}.] The intervention is binary $t \in \{0,1\}$ and assignment status is randomized $T_i \perp {(Y_i(0), Y_i(1))} | X_i$. $\mathbb{P}_Q(T_i = t | X_i = x)$ is known for all $x \in \mathcal{X}$.
    \end{enumerate}
\end{assumption}

Assumption \textbf{BO} (Bounded Outcomes) states that the outcome variable lies in a known interval of length $M$. 
Assumption \textbf{CS} (Covariate Space) states that there are finitely many covariate values and that the probability distribution of covariates is known.
Assumption \textbf{AM} (Assignment Mechanism) states that the intervention is binary (status quo and innovation) and the researcher can conduct a randomized experiment, either with stratified randomization, or with complete randomization, to estimate the optimal assignment rule.
Denote the set of states $Q$ that satisfy Assumption \ref{ass:manski_2004} by $\mathcal{Q}_\mathrm{MK}$.

\begin{remark}
    Note that Assumption \textbf{AM} implies that $\omega$ is a randomized experiment and that $\theta$ is the total (or stratum-specific) sample size. Then, $S^n$ is the experimental sample.
\end{remark}

Let $\mathcal{D}_v=\{d_v:\mathcal{V}\to\mathcal{T}\}$ denote the set of assignment rules measurable with respect to the coarsening $v$.
Using the notation of Section \ref{sec:general_framework}, the set of interventions is binary $\mathcal{T}=\{0,1\}$ and we want to choose an assignment rule $d_v$ so that $W_Q(d)=\mathbb{E}_Q[Y_i(d_v)]$ is maximized. 
The corresponding oracle rule among assignment rules measurable with respect to $v$ is then:
\begin{equation}\label{eq:v_oracle}
    d_v^* = \mathbf{1}\left\{\mathbb{E}_Q[Y_i(1)|v(X_i)=v(x)] > \mathbb{E}_Q[Y_i(0)|v(X_i)=v(x)] \right\}^{|\mathcal{V}|},
\end{equation}
with value
\begin{equation}
    W_Q(d_v^*) = \sum_{\nu \in \mathcal{V}} \mathbb{P}_Q(v(X_i)=\nu)\max_{t \in \{0,1\}}\mathbb{E}_Q[Y_i(t)|v(X_i)=\nu].
\end{equation}
Define
\begin{equation}
    \bar{Y}_{t,x} = \frac{1}{N_{t,x}}\sum_{i=1}^{n} Y_i \cdot \mathbf{1}\{T_i=t\}\cdot \mathbf{1}\{X_i=x\},
\end{equation}
where 
\begin{equation}
    N_{t,x} := \sum_{i=1}^{n}\mathbf{1}\{T_i=t\}\cdot \mathbf{1}\{X_i=x\}.
\end{equation}
For each intervention $t \in \{0,1\}$ and coarse cell $\nu \in \mathcal{V}$, define the sample analog
\begin{equation}
    \bar{Y}_{t,\nu} := \sum_{x \in \mathcal{X}} \bar{Y}_{t,x} \cdot \mathbb{P}_Q(X_i = x|v(X_i)=\nu).
\end{equation}
Then, for a given coarsening $v$ and policy $d_v \in \mathcal{D}_v$, define the estimated welfare
\begin{equation}
    m_\mathrm{CES}(S^n,d_v)
    :=
    \hat{W}(d_v)
    =
    \sum_{\nu \in \mathcal{V}} \mathbb{P}_Q(v(X_i)=\nu) \cdot
    \sum_{t \in \{0,1\}} 
    \mathbf{1}\{d_v(\nu)=t\} \cdot \bar{Y}_{t,\nu}.
\end{equation}
The CES rule based on $v$ can then be written as
\begin{equation}
    \hat{d}_v
    =
    \delta_\mathrm{CES}\{m_\mathrm{CES}(S^n,d_v)\}
    :=
    \arg\max_{d_v \in \mathcal{D}_v}\hat{W}(d_v).
\end{equation}

\begin{remark}
    This aggregate formulation is equivalent to Manski's original value-by-value formulation.
    The reason is that $\hat{W}(d_v)$ is additively separable across cells $\nu \in \mathcal{V}$ and there are no cross-cell constraints linking policy choices across cells.
    Hence the global maximization over $\mathcal{D}_v$ decomposes into $|\mathcal{V}|$ independent comparisons, one for each cell, so equivalently
    \begin{equation}
        \hat{d}_v(\nu)=\mathbf{1}\{\bar{Y}_{1,\nu}>\bar{Y}_{0,\nu}\},
    \end{equation}
    with ties broken in favor of the status quo.
\end{remark}

\begin{remark}
    As already stated in general in Remark \ref{rmk:two_assumptions}, the results in \cite{manski_statistical_2004} are about CES rules. 
    This is equivalent to fixing $(m_\mathrm{CES},\delta_\mathrm{CES})$ and providing theoretical guarantees about this specific pair rather than solving the general problem of Definition \ref{def:policy_choice_problem}.
\end{remark}

\cite{manski_statistical_2004} compares $\hat d_v$ to the oracle rule that conditions on the full covariate vector $X_i$, namely
\begin{equation}\label{eq:x_oracle}
    d_x^*=\mathbf{1}\left\{\mathbb{E}_Q[Y_i(1)|X_i=x] > \mathbb{E}_Q[Y_i(0)|X_i=x]\right\}^{|\mathcal{X}|},
\end{equation}
with value
\begin{equation}
    W_Q(d_x^*) = \sum_{x \in \mathcal{X}} \mathbb{P}_Q(X_i=x)\max_{t \in \{0,1\}}\mathbb{E}_Q[Y_i(t)|X_i=x].
\end{equation}
Instrumental to the comparison with such an oracle, \citet{manski_statistical_2004} defines regret as: 
\begin{equation}
    R_Q(\hat{d}_v)
    =
    \mathbb{E}_{Q_{\omega,\theta}}
    \left[
        W_Q(d_x^*)-W_Q(\hat d_v)
    \right],
\end{equation}
that is, regret relative to the oracle that conditions on the full covariate vector $X_i$. 

Before stating the main result, let for each $\nu \in \mathcal{V}$, 
\begin{equation}
    \mathcal{X}_{\nu}=\{x \in \mathcal{X}: v(x)=\nu\},
\end{equation}
that is, the set of fine covariate values pooled into coarse cell $\nu$. Define
\begin{equation}
    \rho(\nu)
    =
    \max_{A \subseteq \mathcal{X}_{\nu}}
    \min\left\{
        \sum_{x \in A}\mathbb{P}_Q(X_i=x|v(X_i)=\nu),
        1-\sum_{x \in A}\mathbb{P}_Q(X_i=x|v(X_i)=\nu)
    \right\}.
\end{equation}
That is, $\rho(\nu)$ is the largest probability mass that can be placed on the smaller side of a partition of the fine covariate values pooled into cell $\nu$.

\begin{proposition}[Regret bound in \cite{manski_statistical_2004}]\label{prop:bound_mk}
Under Assumption \ref{ass:manski_2004},
\begin{equation}
    \begin{aligned}
        \sup_{Q \in \mathcal{Q}_\mathrm{MK}} R_Q(\hat{d}_v)
        \leq &
        M \sum_{\nu \in \mathcal{V}} \mathbb{P}_Q(v(X_i)=\nu)\rho(\nu) +
        \\ 
        + &
        \frac{M e^{-1/2}}{2}
        \sum_{\nu \in \mathcal{V}} \mathbb{P}_Q(v(X_i)=\nu)
        \left[
            \sum_{x \in \mathcal{X}_{\nu}}
            \mathbb{P}_Q(X_i=x|v(X_i)=\nu)^2
            \left(
                N_{1,x}^{-1}+N_{0,x}^{-1}
            \right)
        \right]^{1/2}.
    \end{aligned}
\end{equation}
\end{proposition}
The first term bounds the difference between the oracles $d^*_x$ (Eq. \ref{eq:x_oracle}) and $d^*_v$ (Eq. \ref{eq:v_oracle}) and can be defined as \textit{approximation} error. 
The second term bounds the difference between the estimated policy $\hat{d}_v$ and the oracle $d^*_v$ and can be defined as \textit{estimation} error.
This terminology was introduced later by \cite{mbakop_model_2021} building an analogy with model selection problems.
Note that, for infinite sample sizes, the estimation error is zero, and for the finest coarsening function (i.e. the identity function), the approximation error is zero.\footnote{The second point follows from the fact that when $v$ is the identity, $\mathcal{X}_\nu=x$ and $\mathbb{P}_Q(X_i = x | X_i = x) = 1$. Therefore, $\rho(\nu) = 0$.} 

\begin{remark}
    Note that in Proposition \ref{prop:bound_mk}, we take the $\sup_{Q \in \mathcal{Q}_\mathrm{MK}}$ on the left-hand side and then keep $Q$-dependent quantities on the right-hand side.
    This is because, under $\mathcal{Q}_\mathrm{MK}$, the covariates' probability law is known and fixed, and therefore results hold pointwise over that fixed law rather than in the worst case over it.
    What is left to vary within $\mathcal{Q}_\mathrm{MK}$ is the true distribution of potential outcomes.
\end{remark}

The approximation error term in Proposition \ref{prop:bound_mk} is sharp. 
As a result, it serves as a lower bound on maximum regret for $\hat{d}_v$: even if we had infinite data, we would incur that welfare loss. 
Moreover, the estimation error term can also be derived for rules that learn from the finest coarsening, that is, the covariate space itself $\hat{d}_x$.
As a result, one can use these two to understand under which conditions on $\theta = (N_{1x},N_{0x})$ it is minimax dominant (see Eq. \ref{eq:minimax_dom} for a definition) to learn $\hat{d}_x$ rather than $\hat{d}_v$.

\begin{proposition}[Sufficient sample sizes]\label{prop:suff_size}
For the full covariate vector $X_i$ to be minimax-dominant relative to the coarsening $v(X_i)$, it is sufficient for $\theta$ to satisfy:
    \begin{equation}
        \sum_{\nu \in \mathcal{V}} \mathbb{P}_Q(v(X_i)=\nu)\rho(\nu) >
        \frac{e^{-1/2}}{2}
        \sum_{x \in \mathcal{X}} \mathbb{P}_Q(X_i=x)
        \left(
            N_{1,x}^{-1}+N_{0,x}^{-1}
        \right)^{1/2}.
    \end{equation}
\end{proposition}
Proposition \ref{prop:suff_size} provides a sufficient condition for $\theta=(N_{1,x},N_{0,x})$ to make it minimax dominant to learn policies in the finest coarsening, i.e. using the full covariate space.

\subsection{Threads following Manski (2004)} \label{sec:unconstrained_ext}
\cite{manski_statistical_2004} opened up a new thread of contributions that studied further the properties of CES rules and the policy choice problem more generally. 

\subsubsection{Local Asymptotics} 
\cite{hirano_asymptotics_2009} formalized the policy choice problem from a local asymptotic perspective using the limit of experiments framework \citep{lecam1986}. 
Relative to \cite{manski_statistical_2004}, the main novelty is that optimality is not studied uniformly over a fixed state space, but locally around \textit{knife-edge} states where the policymaker is nearly indifferent between implementing and not implementing the intervention.

For simplicity, fix a covariate value and suppress it from the notation, as in \cite{hirano_asymptotics_2009}. 
Consider deciding whether or not to implement one binary intervention under full adoption, so $\mathcal{D}=\{0,1\}$. 
Define the welfare contrast
\begin{equation}
    \tau_Q := W_Q(1)-W_Q(0),
\end{equation}
and the corresponding oracle
\begin{equation}
    d_Q^*=\mathbf{1}\{\tau_Q>0\}.
\end{equation}
In the notation of Definition \ref{def:policy_choice_problem}, \cite{hirano_asymptotics_2009} leaves $(\omega,\theta,m)$ abstract, except for requiring that $m(S^n,1)$ be an estimator of $\tau_Q$. 
Writing $\hat{\tau}_n:=m(S^n,1)$, they consider local sequences of DGPs $\{Q_{n,h}\}_{h\in\mathbb{R}}\subseteq\mathcal{Q}$ around a reference state $Q_0$ such that $\tau_{Q_0}=0$ and
\begin{equation}
    \tau_{Q_{n,h}}=\tau_{Q_0}+\frac{h}{\sqrt{n}}=\frac{h}{\sqrt{n}}.
\end{equation}
Hence, $h$ indexes local departures from the knife-edge state in units of $\sqrt{n}$-scaled welfare contrast. 
The asymptotic approximation is then
\begin{equation}
    \sqrt{n}\left(\hat{\tau}_n-\tau_{Q_{n,h}}\right)\rightsquigarrow \mathcal{N}(0,\sigma^2)
\end{equation}
under $Q_{n,h}$. 
Therefore, we can restrict the set of feasible $\Omega \times \Theta_\omega \times \mathcal{M}$ to the one that guarantees this condition to hold.

\cite{hirano_asymptotics_2009} considers a general loss function that can be asymmetric \citep[as defined in][]{tetenov_asymmetric_2012} and evaluated at the mean (see Eq. \ref{eq:bayes}), rather than at the worst case (see Eq. \ref{eq:minimax}).
In this review we consider the case of symmetric regret evaluated at the worst case. Then, we can write 
\begin{equation}
    R_{Q_{n,h}}(\hat d)
    :=
    \mathbb{E}_{\left(Q_{n,h}\right)_{\omega,\theta}}
    \left[
        \tau_{Q_{n,h}}
        \left(
            \mathbf{1}\{\tau_{Q_{n,h}}>0\}-\hat d
        \right)
    \right],
\end{equation}
which coincides with regret, pointwise in the fixed covariate value. 
The main result is that, in the Gaussian limit experiment \citep[as defined in][]{lecam1986}, the minimax rule under symmetric loss is the zero-cutoff rule. 
Mapping it back to the original sample problem yields the plug-in choice rule
\begin{equation}
    \hat d_{\mathrm{HP}}
    :=
    \delta_{\mathrm{HP}}\{m(S^n,1)\}
    =
    \mathbf{1}\{\hat{\tau}_n>0\}.
\end{equation}

\begin{proposition}[Local asymptotic minimax rule in \cite{hirano_asymptotics_2009}]\label{prop:hp09}
Under the regularity conditions above,
\begin{equation}
    \sup_{J\subset\mathbb{R}:|J|<\infty}
    \liminf_{n\to\infty}
    \sup_{h\in J}
    \sqrt{n}\,
    R_{Q_{n,h}}(\hat d_{\mathrm{HP}})
    =
    \inf_{(\delta_n)_{n\geq 1}}
    \sup_{J\subset\mathbb{R}:|J|<\infty}
    \liminf_{n\to\infty}
    \sup_{h\in J}
    \sqrt{n}\,
    R_{Q_{n,h}}(\hat d_n),
\end{equation}
where $\hat d_n:=\delta_n\{m(S,1)\}$, and the infimum is taken over convergent sequences of choice rules with $\delta_n\in\Delta$ for every $n$.
\end{proposition}
Proposition \ref{prop:hp09} states that, once the welfare contrast can be estimated efficiently, no alternative sequence of data-dependent choice rules can uniformly improve on the simple sign rule in a local neighborhood of a knife-edge DGP. 
Unlike Eq. \ref{eq:minimax}, this is a local asymptotic criterion: it compares rules only along sequences of DGPs that approach the knife-edge state at the $1/\sqrt{n}$ rate.

\subsubsection{Exact Finite Sample Behaviour}

\paragraph{Exact Finite Sample Behaviour and \textit{No-Data} Rules.} A complementary contribution by \cite{stoye_minimax_2009} studies the same policy choice problem from the opposite angle: instead of passing to a limit experiment, it derives \textit{exact finite-sample} minimax regret rules. 
For the binary policy choice problem without covariates, binary outcomes $Y_i(t)\in\{0,1\}$, and random assignment in the estimating sample, let
\begin{equation}
    N_t:=\sum_{i=1}^n \mathbf{1}\{T_i=t\},
    \qquad
    \bar Y_t:=\frac{1}{N_t}\sum_{i=1}^n Y_i\mathbf{1}\{T_i=t\},
\end{equation}
with the convention that $N_t(\bar Y_t-1/2)=0$ when $N_t=0$. Then, setting $m(S^n,1)=(N_0,\bar Y_0,N_1,\bar Y_1)$, the exact minimax rule assigns intervention $1$ with probability
\begin{equation}
    \delta_{\mathrm{ST}}\{m(S^n,1)\}
    :=
    \begin{cases}
        1 & \text{if } N_1(\bar Y_1-1/2)-N_0(\bar Y_0-1/2)>0, \\
        1/2 & \text{if } N_1(\bar Y_1-1/2)-N_0(\bar Y_0-1/2)=0, \\
        0 & \text{if } N_1(\bar Y_1-1/2)-N_0(\bar Y_0-1/2)<0.
    \end{cases}
\end{equation}
Under matched pairs (i.e. $N_0 = N_1 = n/2$), this collapses to the empirical success rule with symmetric tie-breaking, while if the status quo is known and only the innovation is sampled, the exact minimax rule becomes a threshold rule in the number of observed successes.

\begin{remark}
    This is the first appearance of fractional rules, which are non-deterministic policy choices. 
    They are non-deterministic as they assign a probability of receiving the intervention, rather than the deterministic choice of assigning the intervention or not.
    As a consequence, $\mathcal{D}$ is implicitly defined as $[0,1]^\ell$ where $\ell$ indexes the number of groups involved in the policy choice.
    The behaviour of such rules was studied in subsequent literature \cite[e.g.][]{stoye_covariates_2012, ishihara_evidence_2026,yata_2025,kitagawa2026,kitagawa_26_bio,Olea_2026}.
    Although fractional rules are proven to be optimal in various contexts, their practical use is very limited as it is hard to imagine a policymaker who assigns an intervention to the full population by random assignment.
\end{remark}

The second main result is about covariates. 
If the state space imposes no cross-covariate restrictions, so that intervention effects can vary arbitrarily across cells of $\mathcal{X}$, then the global finite-sample minimax problem separates into one conditional minimax problem for each covariate value. 
In the notation of this review, the minimax rule takes the form $\hat d(x)=\delta_x^*\{m(S_x^n,1)\}$, where $S_x^n$ denotes the subsample with $X_i=x$ and each $\delta_x^*$ is minimax for the corresponding conditional problem. 
This result sharpens \cite{manski_statistical_2004}: absent restrictions on the intervention effect function over covariates, pooling information across covariate values (e.g. via the coarsening map $v(x)$ defined above) is never minimax optimal. 
The most striking case is when $X$ is continuously distributed. Then, for any fixed target value $x$, a finite sample contains no repeated observations with exactly that covariate value with probability one, so the conditional subsample $S_x^n$ is generically empty. 
If the outcomes under both the status quo and the intervention are unknown, the exact minimax rule assigns the intervention with probability $1/2$ for every $x$, independently of the realized data. 
If instead the status quo welfare is known, the minimax rule is still a no-data rule, but it assigns intervention $1$ with probability $1-W_Q(0)$, where $W_Q(0)$ is conditional on $x$, at each covariate value. 
The intuition is that, without smoothness restrictions on the potential outcomes function, observations with similar covariate values cannot be used to learn about the welfare contrast at $x$ uniformly over the state space. 
Hence, with continuous covariates and a sufficiently rich $\mathcal{Q}$, finite-sample minimax regret may force the researcher to ignore the sample altogether, even though such rules can be pointwise dominated once stronger structure is imposed. 
\cite{stoye_covariates_2012} refines this result and shows that if the state space is restricted so that covariates have bounded influence on welfare, for example
\begin{equation}
    |W_Q(t|x)-W_Q(t|x')|\leq \kappa
    \qquad
    \forall t\in\{0,1\},\ x,x'\in\mathcal{X},
\end{equation}
then there exists a strictly positive region in which the exact minimax rule pools information across covariate values.
Thus, pooling information across covariate values can be minimax optimal when their importance is bounded. 

\paragraph{Shrinkage for Policy Choice.} The value of using covariates is revisited again by \cite{Ishihara_2025}, who focus on the intermediate region between complete pooling and cell-by-cell empirical success rules. 
Working with finitely many covariate cells, Gaussian approximations for cell-specific welfare-contrast estimators, and assuming $|\tau_Q(x)-\bar{\tau}_Q|\leq \kappa$ (or equivalently that the intervention effect function has a bounded second moment), they propose shrinkage rules that replace each raw estimate with a convex combination of the cell-specific estimate and the pooled estimate. 
The shrinkage parameter is chosen by minimizing a tractable upper bound on maximum regret, so the resulting rule interpolates between CES and full pooling. 
The main result says that when heterogeneity in intervention effects across covariate values is neither negligible nor unbounded, shrinkage can strictly improve on both extremes in worst-case regret terms.

A related use of shrinkage appears in \cite{moon2026}, who studies a different policy problem in which the planner chooses local changes across many policies rather than a binary assignment rule for one policy. 
For each policy $j$, the planner faces noisy estimates of benefit and net cost and wishes to choose a vector of small policy changes that maximizes expected welfare. 
The oracle rule depends on posterior mean benefit and posterior mean net cost for each policy. 
\cite{moon2026} proposes an empirical Bayes approach that shrinks policy-specific estimates toward a common distribution learned from the full collection of policies and shows that this approach can approximate the oracle allocation, while a raw plug-in rule may fail. 
Relative to \cite{Ishihara_2025}, the paper therefore applies the same broad shrinkage logic to a different dimension of the problem: pooling information across policies rather than across covariate cells.

The same broad lesson appears in \cite{yamin2026}, who studies poverty targeting when the policymaker must allocate nonnegative transfers across households subject to a fixed budget. 
The planner does not observe true poverty gaps, but only noisy estimates of them, so the policy problem is how to translate noisy signals into a feasible transfer allocation. 
The paper shows that the plug-in rule that treats estimated poverty gaps as true is inadmissible on a compact parameter space.
This implies there exists some feasible rule that weakly lowers regret uniformly in the state space and strictly lowers it in some cases.
They propose a nonparametric empirical Bayes rule that replaces each poverty estimate with its posterior mean and then applies the same constrained allocation logic as under full information.
Finally, they characterize the oracle Bayes allocation and derive regret guarantees relative to that oracle.
This procedure is shown to have sizeable gains relative to the plug-in rule in simulations.
Relative to \cite{moon2026}, the shrinkage now pools information across households rather than across policies, but the message is similar: posterior shrinkage can improve downstream policy allocation when the underlying signals are noisy.

\subsubsection{Different Loss Functions}
\paragraph{Asymmetric Regret.} \cite{tetenov_asymmetric_2012} extends \cite{manski_statistical_2004} to allow for planners who are differently averse to type I and type II errors. 
Keep the one-intervention, full-adoption problem with known status quo welfare $W_Q(0)$, innovation welfare $W_Q(1)$ learned from a finite sample, and welfare contrast $\tau_Q:=W_Q(1)-W_Q(0)$. 
Let $\hat\tau_n:=m(S^n,1)$ denote an estimator of $\tau_Q$.
For a threshold rule $\hat d_T:=\mathbf{1}\{\hat\tau_n>T\}$, regret takes the statewise form
\begin{equation}
    R_Q(\hat d_T)
    =
    \begin{cases}
        -\tau_Q \cdot \mathbb{P}_Q(\hat\tau_n>T) & \text{if } \tau_Q\leq 0,\\
        \tau_Q \cdot \mathbb{P}_Q(\hat\tau_n\leq T) & \text{if } \tau_Q>0.
    \end{cases}
\end{equation}
The first line is the welfare loss from adopting an inferior innovation, while the second is the welfare loss from rejecting a superior one.
This motivates defining, for a generic data-dependent rule $\hat d$,
\begin{equation}
    \bar R^{\mathrm{I}}(\hat d)
    :=
    \sup_{Q\in\mathcal{Q}:\tau_Q\leq 0} R_Q(\hat d),
    \qquad
    \bar R^{\mathrm{II}}(\hat d)
    :=
    \sup_{Q\in\mathcal{Q}:\tau_Q>0} R_Q(\hat d).
\end{equation}
Therefore $\bar R^{\mathrm{I}}(\hat d)$ is maximum Type I regret and $\bar R^{\mathrm{II}}(\hat d)$ is maximum Type II regret. 
\citet{tetenov_asymmetric_2012} studies the weighted minimax problem
\begin{equation}
    \delta_K
    \in
    \argmin_{\delta\in\Delta}
    \max\left\{
        K\bar R^{\mathrm{I}}(\hat d),
        \bar R^{\mathrm{II}}(\hat d)
    \right\},
    \qquad
    \hat d:=\delta\{m(S^n,1)\},
\end{equation}
with $K>0$.
The symmetric minimax criterion is the special case $K=1$, which recovers the cases studied in \cite{manski_statistical_2004}, \cite{stoye_minimax_2009}, \cite{stoye_covariates_2012}.

\citet{tetenov_asymmetric_2012} shows that in the Gaussian experiment where $\hat\tau_n\sim\mathcal{N}(\tau_Q,\sigma^2)$, it is sufficient to search for the minimax choice rule among threshold rules of the form $\hat d_T=\mathbf{1}\{\hat\tau_n>T\}$. 
Hence, although the original choice variable is the decision rule $\delta\in\Delta$, after this reduction the problem can be re-written as the choice of a scalar threshold $T\in\mathbb{R}$.
Using the previous display and writing $h:=\tau_Q/\sigma$, their maximum Type I and Type II regrets can be written as
\begin{equation}
    \bar R^{\mathrm{I}}(\hat d_T)
    =
    \sigma\sup_{h\leq 0}\{-h\Phi(h-T/\sigma)\},
    \qquad
    \bar R^{\mathrm{II}}(\hat d_T)
    =
    \sigma\sup_{h>0}\{h\Phi(T/\sigma-h)\}.
\end{equation}
Therefore the asymmetric minimax rule is equivalently characterized by the threshold choice
\begin{equation}
    T_K
    \in
    \arg\min_{t\in\mathbb{R}}
    \max\left\{
        K\sup_{h\leq 0}\{-h\Phi(h-t)\},
        \sup_{h>0}\{h\Phi(t-h)\}
    \right\},
\end{equation}
where $T_K$ denotes the unique minimizer in the normalized problem with $\sigma=1$.
Equivalently, $T_K$ is the unique normalized cutoff satisfying
\begin{equation}
    K\sup_{h\leq 0}\{-h\Phi(h-T_K)\}
    =
    \sup_{h>0}\{h\Phi(T_K-h)\}.
\end{equation}
The corresponding threshold in the original Gaussian problem is $T=\sigma T_K$.
Hence the optimal threshold moves to the right with $K$: compared to symmetric minimax, the planner requires stronger evidence before recommending the innovation. 
For Bernoulli outcomes with known status quo mean $p_0$, the exact finite-sample solution has the same structure: it is a threshold rule in the number of observed successes, defined by the unique cutoff that equalizes weighted maximum Type I and Type II regret. 
Its large-sample approximation is
\begin{equation}
    \hat d_{K,\mathrm{N}}
    =
    \mathbf{1}\left\{
        \bar Y_1-p_0
        >
        \frac{\sigma_0 T_K}{\sqrt{n}}
    \right\},
    \qquad
    \sigma_0^2:=p_0(1-p_0),
\end{equation}
which collapses to the plug-in rule when $K=1$. 
An especially insightful interpretation is that one-sided hypothesis-test rules correspond to asymmetric minimax rules for particular values of $K$: in the normal model, a level $\alpha=0.05$ rule is equivalent to $K\approx 102$, while $\alpha=0.01$ corresponds to $K\approx 970$. 
This makes precise how strongly conventional testing procedures privilege avoiding mistaken adoption relative to missing a beneficial innovation.

\paragraph{Non-linear Regret.}
\cite{kitagawa_26_bio} consider the policy choice problem of \citet{manski_statistical_2004} under non-linear regret. 
For a possibly fractional rule $\hat d:=\delta\{m(S^n,1)\}\in[0,1]$, they evaluate choice rules through the nonlinear regret risk
\begin{equation}
    r_Q(\hat d)
    :=
    \mathbb{E}_{Q_{\omega,\theta}}
    \left[
        g\left(
        R(\hat{d})
        \right)
    \right],
\end{equation}
where $g:\mathbb{R}_+\to\mathbb{R}_+$ is increasing and nonlinear and 
\begin{equation}
    R(\hat{d}) := W_Q(d^*) - W_Q(\hat{d}),
\end{equation} 
that is, regret on the sample that realized, not in expectation over the sampling distribution (which is denoted with $R_Q(\hat{d})$).
Their benchmark case is mean-square regret, $g(r)=r^2$, for which
\begin{equation}
    r_Q(\hat d)
    =
    R_Q(\hat d)^2
    +
    \mathrm{Var}_{Q_{\omega,\theta}}
    \left[
        R(\hat{d})
    \right],
\end{equation}
so the planner penalizes not only average regret but also the volatility of regret induced by sampling uncertainty. 
This change has a first-order implication: singleton rules are no longer essentially complete (in the sense that restricting the policy space to such rules is without loss), so optimal rules are generally fractional. 
In the Gaussian limit experiment \citep[as defined in][]{hirano_asymptotics_2009} where $\hat\tau_n:=m(S^n,1)\sim\mathcal{N}(\tau_Q,\sigma^2)$, the finite-sample minimax mean-square regret rule takes the simple logistic form
\begin{equation}
    \hat d_{\mathrm{KLQ}}
    :=
    \delta_{\mathrm{KLQ}}\{m(S^n,1)\}
    =
    \frac{
        \exp\left(2\tau^*\hat\tau_n/\sigma\right)
    }{
        1+\exp\left(2\tau^*\hat\tau_n/\sigma\right)
    },
    \qquad
    \tau^*\approx 1.23,
\end{equation}
and the same structure survives in regular parametric models as a logistic transform of the $t$-statistic. 

\paragraph{Certified Decisions.}
\cite{andrews_chen_2025} study how inference can be turned into decision recommendations with explicit guarantees. 
In the notation of this review, suppose the researcher constructs a confidence set $\hat{\mathcal{Q}}_n\subseteq\mathcal{Q}$ for the unknown state $Q$. 
They then consider \textit{as-if} decisions of the form
\begin{equation}
    \hat d_{\mathrm{AC}}
    \in
    \arg\max_{d\in\mathcal{D}}
    \inf_{Q'\in\hat{\mathcal{Q}}_n}W_{Q'}(d),
    \qquad
    \hat r_n
    :=
    \sup_{Q'\in\hat{\mathcal{Q}}_n}
    \left[
        W_{Q'}(d_{Q'}^*)-W_{Q'}(\hat d_{\mathrm{AC}})
    \right],
\end{equation}
where $d_{Q'}^*:=\arg\max_{d\in\mathcal{D}}W_{Q'}(d)$. 
Thus $\hat r_n$ is a high-probability upper bound on the regret of the recommended decision. 
If $\hat{\mathcal{Q}}_n$ covers the true state with probability $1-\alpha$, then
\begin{equation}
    \mathbb{P}\left\{
        W_Q(d^*)-W_Q(\hat d_{\mathrm{AC}})
        \leq
        \hat r_n
    \right\}
    \geq
    1-\alpha.
\end{equation}
Their main result is that, among rules paired with such high-probability regret bounds, there is essentially no loss in restricting attention to these confidence-set-based decisions. 
The paper therefore gives a general foundation for using inferential objects to support ambiguity-averse downstream decision-makers.

\paragraph{Policy Learning with Confidence.}
\cite{chernozhukov_lee_rosen_sun_2026} provide a concrete application of this logic to policy choice when, for each feasible rule $d\in\mathcal{D}$, the researcher has an estimate $\hat W_n(d)$ of welfare together with a standard error $\hat s_n(d)$. 
Rather than choosing the rule with the largest estimated welfare, they consider \textit{risk-aware} rules that trade off performance and estimation risk and therefore move along an efficient frontier of feasible welfare-risk pairs. 
Their proposed PoLeCe rule chooses
\begin{equation}
    \hat d_{\mathrm{PLC}}
    \in
    \arg\max_{d\in\mathcal{D}}
    \left\{
        \hat W_n(d)-\hat c_{1-\alpha}\hat s_n(d)
    \right\},
\end{equation}
where $\hat c_{1-\alpha}$ is calibrated so that the objective is a one-sided lower confidence bound for welfare. 
Hence the rule selects the policy with the highest guaranteed welfare at confidence level $1-\alpha$, rather than the policy with the highest point estimate. 
Relative to the minimax contributions above, the paper does not modify the loss function directly; instead it changes the decision criterion so that sampling uncertainty enters policy choice through a reporting guarantee.

\subsubsection{Partial Identification}
The conceptual distinction between making decisions under \textit{uncertainty}, i.e. the problem of solving a population-wide problem with data that provide point identification, and making decisions under \textit{ambiguity}, i.e. solving a population problem where data only provide partial identification in the sense of \cite{manski2003partial}, was first noted by \cite{MANSKI2000415}.
Up to this point, we focused on the first kind of problem, but a thread following \cite{MANSKI2000415, manski_statistical_2004} has considered settings where both sources of difficulty arise.

\paragraph{Univariate Partial Identification.} A first contribution in this direction is \cite{stoye_covariates_2012}, who studies the binary full-adoption problem when the experiment does not point-identify the target-population welfare contrast $\tau_Q:=W_Q(1)-W_Q(0)$, but only an estimating-population contrast, say $\tau_Q^{S}$, linked to $\tau_Q$ by
\begin{equation}
    \tau_Q \in [a\tau_Q^{S}-b,a\tau_Q^{S}+b]\cap[-1,1],
\end{equation}
with $a\in(0,1]$ and $b>0$ summarizing the severity of the failure of internal or external validity. 
This representation nests, among others, selective noncompliance and selective sampling.
In the Gaussian experiment, letting $\hat\tau_n:=m(S^n,1)\sim \mathcal{N}(\tau_Q^{S},\sigma^2)$, \cite{stoye_covariates_2012} shows that the minimax regret rule is
\begin{equation}
    \hat d_{\mathrm{ST}}
    =
    \delta_{\mathrm{ST}}\{m(S^n,1)\}
    :=
    \begin{cases}
        \mathbf{1}\{\hat\tau_n>0\} & \text{if } b\leq (\pi/2)^{1/2}\sigma a,\\
        \Phi\left(\hat\tau_n;0,\frac{2}{\pi}(b/a)^2-\sigma^2\right) & \text{if } (\pi/2)^{1/2}\sigma a<b<1,\\
        1/2 & \text{if } b\geq 1,
    \end{cases}
\end{equation}
In the middle case, $\Phi(x;0,v)$ denotes the c.d.f. of a centered Gaussian variable with variance $v$ evaluated at $x$, so the rule assigns the intervention with probability $\Phi(\hat\tau_n;0,(2/\pi)(b/a)^2-\sigma^2)$. 
Equivalently, the planner adds a centered Gaussian noise with variance $(2/\pi)(b/a)^2-\sigma^2$ to the experimental signal and then applies the sign rule to the perturbed signal. 
The condition $(\pi/2)^{1/2}\sigma a<b<1$ indexes an intermediate region: ambiguity is already too large for the pure plug-in rule $\mathbf{1}\{\hat\tau_n>0\}$ to remain minimax-optimal, but not yet so large that it is optimal to ignore the data altogether and set $\hat d_{\mathrm{ST}}=1/2$.
Hence, when ambiguity is small relative to sampling precision, the planner should follow the sign of the experimental signal exactly as in the point-identified benchmark; once the wedge between $\tau_Q$ and $\tau_Q^{S}$ becomes large enough, minimax regret optimally attenuates the signal through randomization, and for sufficiently severe validity failures collapses to the no-data rule.

\paragraph{General Partial Identification.} 
\cite{yata_2025} shows how to solve finite-sample minimax regret problems in a more general partially identified environment. 
The researcher observes a Gaussian statistic $Z^n\sim \mathcal{N}(g(\vartheta),\sigma^2 I_n)$, where $\vartheta$ may be high-dimensional or infinite-dimensional, $g:\mathcal{H}\to\mathbb{R}^n$ is linear, and $\mathcal{H}$ is a nonempty, convex, and centrosymmetric parameter space. 
Welfare is summarized by the linear contrast $\Lambda(\vartheta)=W_1(\vartheta)-W_0(\vartheta)$. 
For a fixed reduced-form value $\mu$, the identified set
\begin{equation}
    I(\mu):=\{\Lambda(\vartheta):g(\vartheta)=\mu,\ \vartheta\in\mathcal{H}\},
\end{equation}
collects all welfare contrasts that remain feasible once the observables are fixed and identify $\mu$. 
Hence partial identification arises whenever $I(\mu)$ contains both positive and negative values: the data identify the reduced form, but not whether introducing the intervention maximizes welfare. 
This setup nests the case considered by \cite{stoye_covariates_2012}, as well as other partial-identification problems that are multivariate in nature, such as evidence aggregation for policy choice \citep[as defined in][]{ishihara_evidence_2026}.

The first-order contribution of the paper is the theory used to solve this problem \textit{exactly} in finite samples. 
For each centrosymmetric line $[-\bar{\vartheta},\bar{\vartheta}]\subseteq\mathcal{H}$, Yata first shows that the minimax rule is the sign rule $\mathbf{1}\{g(\bar{\vartheta})'Z^n\geq 0\}$ when the sample is informative on that line, and a $1/2$-randomization rule when it is not. 
Then, he defines
\begin{equation}
    \psi(\epsilon):=\sup\{\Lambda(\vartheta):\|g(\vartheta)\|\leq \epsilon,\ \vartheta\in\mathcal{H}\},
\end{equation}
which measures the largest welfare contrast compatible with signal strength $\epsilon$. 
The hardest one-dimensional subproblem is then indexed by
\begin{equation}
    \epsilon^* \in \arg\max_{\epsilon\in[0,\tau^*\sigma]} \psi(\epsilon)\Phi(-\epsilon/\sigma),
    \qquad
    \tau^* \in \arg\max_{t\geq 0} t\Phi(-t).
\end{equation}
Thus, Nature chooses the signal strength that maximizes welfare contrast times error probability. 
Let $\vartheta_{\epsilon^*}$ attain the modulus at $\epsilon^*$, and $w^*$ denote the local direction along which the upper bound $\bar I(\mu):=\sup I(\mu)$ expands fastest at $\mu=0$. 
Then, \cite{yata_2025} shows that a minimax regret rule depends only on one scalar least-favorable index:
\begin{equation}
    \hat d_{\mathrm{YT}}
    =
    \delta_{\mathrm{YT}}\{Z^n\}
    :=
    \begin{cases}
        \mathbf{1}\{g(\vartheta_{\epsilon^*})'Z^n\geq 0\}
        & \text{if } 2\phi(0)\psi(0)/\psi'(0)<\sigma,\\
        \Phi\left(
        \frac{(w^*)'Z^n}{\sqrt{(2\phi(0)\psi(0)/\psi'(0))^2-\sigma^2}}
        \right)
        & \text{if } 2\phi(0)\psi(0)/\psi'(0)>\sigma,\\
        1/2
        & \text{if } \psi'(0)=0,
    \end{cases}
\end{equation}
where $\phi$ and $\Phi$ denote the standard normal p.d.f. and c.d.f., respectively.\footnote{There is a knife-edge threshold rule when $2\phi(0)\psi(0)/\psi'(0)=\sigma$, and the minimax risk is $R(\mathcal{H})=\psi(\epsilon^*)\Phi(-\epsilon^*/\sigma)$. }
The key intuition is simple: once the least favorable one-dimensional subproblem is found, the original multivariate problem collapses to deciding on the sign of a single index, possibly after adding noise when local ambiguity is too severe. 

\paragraph{Local Asymptotic Partial Identification.}
A complementary route is taken by \cite{christensen_payoffs_2025}, who study a closely related binary choice problem under partial identification, but from the local asymptotic perspective of \cite{hirano_asymptotics_2009}. 
The researcher observes data $X^n\sim P_{n,\mu}$ informative about a point-identified reduced-form parameter $\mu\in\mathcal{M}$, while policy payoffs depend on a structural object $\vartheta\in\Theta_0(\mu)$ that remains only set identified conditional on $\mu$. 
For each action $d\in\mathcal{D}$, they define the conditional worst-case risk
\begin{equation}
    R(d,\mu):=\sup_{\vartheta\in\Theta_0(\mu)} r(d,\vartheta,\mu),
\end{equation}
and the corresponding oracle
\begin{equation}
    \delta^o(\mu)\in\arg\min_{d\in\mathcal{D}}R(d,\mu),
\end{equation}
which is infeasible because $\mu$ must be estimated. 
Localizing around a reference value $\mu_0$ through $\mu=\mu_0+h/\sqrt{n}$, they rank sequences of rules $(\delta_n)_{n\geq 1}$ by the integrated local excess risk
\begin{equation}
    \mathcal{R}(\{\delta_n\};\mu_0)
    :=
    \int
    \limsup_{n\to\infty}
    \sqrt{n}
    \left[
        \mathbb{E}_{\mu_0+h/\sqrt{n}}
        R\left(
        \delta_n(X^n),
        \mu_0+\frac{h}{\sqrt{n}}
        \right)
        -
        \min_{d\in\mathcal{D}}
        R\left(
        d,
        \mu_0+\frac{h}{\sqrt{n}}
        \right)
    \right]
    dh.
\end{equation}
If $\pi_n(\cdot)=\pi(\cdot\mid X^n)$ denotes a posterior, or quasi-posterior, for $\mu$, the main result is that the feasible rule
\begin{equation}
    \hat d_{\mathrm{CMS}}
    :=
    \delta_{\mathrm{CMS}}\{X^n\}
    \in
    \arg\min_{d\in\mathcal{D}}
    \int R(d,\mu)\,d\pi_n(\mu)
\end{equation}
is asymptotically optimal under this criterion, and so is any bootstrap or quasi-Bayes implementation that is asymptotically equivalent to it. 
The key technical point is that partial identification typically makes $R(d,\mu)$ only directionally differentiable in $\mu$, so the usual plug-in rule $\delta^o(\hat\mu_n)$ need not be asymptotically equivalent to $\hat d_{\mathrm{CMS}}$ and may therefore be suboptimal. 
Hence, relative to \cite{yata_2025}, the paper does not solve the finite-sample minimax regret problem directly; rather, it extends the Hirano-Porter local asymptotic framework to environments in which ambiguity is first profiled through $\Theta_0(\mu)$ and then integrated over local uncertainty in $\mu$.

\paragraph{Multiplicity and Least Randomization.}
\cite{Olea_2026} revisit a closely related class of partially identified Gaussian problems from a different angle. 
Rather than constructing a minimax regret rule, their focus is on the structure of the minimax regret solution set. 
Their main message is that minimax regret may be highly non-unique: in general, there exist many minimax-regret-optimal rules, and fractional rules arise as a general feature of the minimax regret criterion itself. 
To recover uniqueness of the minimax rule, they propose a refinement based on least randomization, characterizing the minimax regret rule that randomizes on the smallest set of data realizations. 
In this sense, the paper is complementary to \cite{yata_2025}: Yata proves existence and derives a formula for a minimax regret rule based on a scalar least-favorable index, while \cite{Olea_2026} show that, once the problem is in the randomized region, such rules need not be unique and may be ranked further using least-randomization as a criterion.

\paragraph{External Validity as Distributional Robustness.}
A conceptually distinct contribution is \cite{adjaho_external_2022}, who study policy choice when the target population may differ from the experimental one. 
Let $P$ denote the experimental distribution of $(X,Y_0,Y_1)$ and let $d:\mathcal{X}\to\{0,1\}$ be a policy rule. 
Instead of maximizing experimental-population welfare
\begin{equation}
    W_P(d):=\mathbb{E}_P[Y_1d(X)+Y_0(1-d(X))],
\end{equation}
they propose the robust welfare criterion
\begin{equation}
    RW_\epsilon(d):=\inf_{Q:W(P,Q)\leq \epsilon}\mathbb{E}_Q[Y_1d(X)+Y_0(1-d(X))],
\end{equation}
where $W(P,Q)$ is a Wasserstein distance and $\epsilon$ has the interpretable meaning of the largest admissible shift in the average intervention effect between the experimental and target populations. 
Their sharpest result concerns shifts in potential outcomes only, holding the covariate distribution fixed: in that case,
\begin{equation}
    RW_\epsilon(d)=W_P(d)-\epsilon
\end{equation}
when outcomes are unbounded below, up to the obvious lower-support truncation when they are bounded. 
Hence the ordering of rules is unchanged, so any policy rule that is optimal, or has small regret, in the experimental population remains optimal, or has the same regret guarantee, under this notion of external validity. 
A related extension allows the covariate distribution to shift from density $p$ in the estimating population to density $q$ in the target population. 
Writing $\rho(x):=q(x)/p(x)$, the relevant benchmark becomes the reweighted welfare $\mathbb{E}_P[(Y_1d(X)+Y_0(1-d(X)))\rho(X)]$, and the same invariance result holds: once outcome drift is modeled through a Wasserstein neighborhood, the robust counterpart preserves the ranking of rules up to the same $\epsilon$ penalty. 
If instead both covariates and potential outcomes are allowed to shift adversarially in unknown ways, robust welfare depends on the distance of each covariate value from the intervention status dictated by the rule.

A closely related contribution is \cite{kido_distributionally_2022}, who studies policy choice when the source and target populations may differ not only in covariate composition but also in the conditional distribution of potential outcomes. 
Relative to \cite{adjaho_external_2022}, the paper adopts a different Wasserstein ambiguity set, centered pointwise on the source conditional distribution and combined with knowledge of the target covariate distribution. 
This leads to a more constructive policy-learning message: robustness can be implemented by evaluating each rule under a systematically pessimistic version of the estimating population payoffs. 
Under suitable conditions, the resulting distributionally robust rule coincides with the naive reweighted rule, but in general the ranking of policies can change, so external-validity concerns may alter the optimal policy rather than simply add a common penalty \citep[as in][]{adjaho_external_2022}. 
The paper also derives regret guarantees for the estimated robust rule.

\paragraph{Geometric Control under Donor-Target Mismatch.} \cite{opocher_geom_2026} studies a setting in which the estimating sample comes from an innovated donor population, while the policy must be applied to a distinct target population. 
Relative to the previous literature on external validity and partial identification, the paper fixes a threshold decision rule and focuses on the choice of estimator $m$. 
It introduces certification: an estimator yields certified decisions if, whenever $|\tau(x)|$ is sufficiently large, the probability of recommending the wrong sign is uniformly controlled. 
The paper shows that certification implies a bound on worst-case compensation loss. 
Restricting attention to matching estimators with positive weights, it then derives a certification radius composed of a geometric approximation term and stochastic terms, and shows that in a large-sample regime the problem becomes purely geometric. 
Within this class, the Delaunay interpolant delivers the best asymptotic affordability guarantee. 
This geometric perspective allows one to identify the exact point in the covariate space that maximizes the worst-case loss and therefore guides data collection processes capable of bringing the loss down to a target level in as few collection steps as possible.

\subsubsection{Adaptive Collection Designs}
\cite{kato_adaptive_2025} shifts attention from the terminal policy rule to the experimental design itself. 
He studies the binary policy choice problem under full adoption when the experimenter can adaptively update intervention assignment probabilities during data collection, rather than fixing them ex-ante.
In the notation of Definition \ref{def:policy_choice_problem}, this is a problem where the main design object is the data-collection process $\omega$, while the terminal rule $\delta$ remains the simple recommendation of the intervention with the highest estimated mean outcome.

Formally, the experiment lasts for $T$ rounds. At each round $t$, the experimenter chooses an intervention $D_t\in\{0,1\}$ based on past observations $\{(D_s,Y_s)\}_{s=1}^{t-1}$ and then, after the allocation phase, recommends
\begin{equation}
    d_\mu^*=\arg\max_{d\in\{0,1\}}\mu_d,
\end{equation}
where $\mu_d:=\mathbb{E}[Y_d]$.
Performance is evaluated by simple regret,
\begin{equation}
    \mathrm{Regret}_\mu(\hat d):=\mathbb{E}_\mu[\mu_{d_\mu^*}-\mu_{\hat d}],
\end{equation}
that is, the welfare loss from recommending the wrong intervention after the adaptive experiment.

The proposed design is a two-stage Neyman allocation. 
The first stage allocates both interventions uniformly to estimate their standard deviations.
The second stage allocates intervention $d$ proportionally to the estimated standard deviation $\sigma_d$.
At the end of the experiment, the experimenter recommends the intervention with the larger sample mean.
The main result is that, over a class of mean-parameterized canonical exponential families, this adaptive design is both minimax and Bayes optimal in the sense that its regret upper bounds match the corresponding lower bounds exactly.
In particular, if $\bar{\sigma}_d:=\sup_{\mu\in\mathcal{M}}\sigma_d(\mu)$, then
\begin{equation}
    \limsup_{T\to\infty}\sqrt{T}\sup_{\mu\in\mathcal{M}^2}\mathrm{Regret}_\mu(\hat d)
    \leq
    (\bar{\sigma}_1+\bar{\sigma}_0)\Phi(-1),
\end{equation}
and \cite{kato_adaptive_2025} proves a matching minimax lower bound.
Thus, relative to \cite{manski_statistical_2004}, the paper endogenizes sequential data collection and shows that adaptive Neyman allocation is first-best when the researcher's goal is an ex-post assignment rule rather than average effect estimation.
\subsection{Constrained Policy Spaces}\label{sec:constrained}
A second wave of research started from the observation that restrictions on the complexity of the policy space can be leveraged together with restrictions on the state space to control worst-case regret.

This fact was first noted by \cite{kitagawa_who_2018}.
They introduce Empirical Welfare Maximization (EWM), a pair $(m, \delta)$ that fixes $m$ to the empirical average of welfare produced by an intervention and sets $\delta$ to select the maximum across the policy space. 
The main theoretical contribution lies in proving that one can leverage (i) a constraint on the complexity of the policy space and (ii) a \textit{within} policy space definition of regret to derive rate-sharp regret bounds.
Moreover, EWM is shown to be minimax-rate optimal (see Eq. \ref{eq:minimax_rate_opt} for a definition).
The mechanics of the main results build on well-known results in statistical learning \citep[see, e.g.,][]{van_der_vaart_weak_2023}.
This powerful connection renewed the interest in the policy learning problem and empowered a sizeable second wave of theoretical research that adapted the general results in specific settings.

The metric of complexity considered is the VC-dimension \citep[see, e.g.,][for a definition]{van_der_vaart_weak_2023}. 
In particular, \cite{kitagawa_who_2018} consider collections of assignment rules $\mathcal{D}_v = \{d : \mathcal{X} \to \{0,1\}\}$ such that $: \mathrm{VC}(\mathcal{D}_v)=v$.
In general, the VC dimension measures the number of points a function can \textit{shatter}.
In the context of assignment rules (i.e. functions that determine \textit{who} should be assigned to an intervention), the VC dimension counts how many distinct points of $\mathcal{X}$ can be labeled as either assigned or not while respecting the rule.
Intuitively, this measure bounds how flexible or complex an assignment rule can be, and therefore how hard the statistical problem of finding the true optimum within that space is.

By \textit{within} policy class regret we mean that the oracle is defined as:
\begin{equation}\label{eq:d_v}
    d^* = \arg\max_{d \in \mathcal{D}_v} W_Q(d),
\end{equation}
rather than
\begin{equation}\label{eq:d_inf}
    d^* = \arg\max_{d \in \cup_{v=1}^\infty \mathcal{D}_v} W_Q(d).
\end{equation}

Therefore, the performance of any choice of $(m, \delta)$ is evaluated relative to an oracle that is imposed to make choices just as complex as the policy space. 
As a result, the first-best policy choice $d^{\mathrm{FB}}(x) = \mathbf{1}\{\tau(x)>0\}$ may not be available to the oracle, unless $d^{\mathrm{FB}}\in \mathcal{D}_v$. 
If that is not the case, the oracle would instead need to choose the best approximation of $d^\mathrm{FB}$ given the policy class.

\begin{remark}
    Note that Equations \ref{eq:d_v} and \ref{eq:d_inf} highlight the fact that the oracle also makes its own choices of $\delta$ and that this formulation constrains $\delta$ such that the oracle optimum is searched only within $\mathcal{D}_v$.
    This is a key element of the main results in \cite{kitagawa_who_2018} and raises the more conceptual question of what the oracle should or should not be allowed to do, given the statistical decision problem of interest.
    One way to navigate this question is to ask the researcher: if you had infinite resources, what would you do? In this specific case, if the answer is `I would still choose within $\mathcal{D}_v$ because the PM only allows this class', then the results in \cite{kitagawa_who_2018} apply.
    However, if the PM would instead search for the first-best policy given the infinite budget, then the oracle should be defined differently, for instance as in \cite{manski_statistical_2004}.
\end{remark}

\begin{remark}
    The choice of $\mathcal{D}_v$ is left as an exogenous choice of the policymaker in the setting of \cite{kitagawa_who_2018}.
    As a result, the theoretical guarantees hold pointwise for any given choice of $v$.
\end{remark}

\cite{kitagawa_who_2018} place the following restrictions on the state space.

\begin{assumption}[\cite{kitagawa_who_2018}'s state space]\label{ass:kt_18} $ $
    \begin{itemize} 
        \item[\textbf{AM}.] The intervention is binary $t\in \{0,1\}$ and the assignment status is randomized given covariates $(Y_i(1),Y_i(0)) \perp T_i | X_i$.
        \item[\textbf{BO}.] Outcomes satisfy: $\exists\ M < \infty$: $|Y_i(t)| \leq M/2$ for all $t \in \mathcal{T}$.
        \item[\textbf{SO}.] Each unit has positive probability of being assigned to either counterfactual state $\exists\ \kappa \in (0,1/2)$: the propensity score $e(x)$ satisfies $e(x) \in [\kappa, 1-\kappa]$ $\forall x \in \mathcal{X}$. 
        \item[\textbf{VC}.] VC-Class: the class of decision sets has finite VC-dimension $v<\infty$ and is countable.
    \end{itemize}
\end{assumption}

Assumption \textbf{AM} (Assignment Mechanism) characterizes a quasi-experimental environment in which intervention assignment is independent of potential outcomes conditional on covariates, and the potential outcome of each unit $i$ depends only on their own intervention status. 
Assumption \textbf{BO} (Bounded Outcomes) implies that both potential outcomes, and thus intervention effects, are uniformly bounded in absolute value by $M/2$.
Boundedness is a standard condition in the statistical learning literature as it enables the use of uniform concentration inequalities \citep[see, e.g.][]{hoeffding_probability_1963,van_der_vaart_weak_2023}. 
Assumption \textbf{SO} (Strict Overlap) is standard in the causal inference literature and guarantees that all units have a positive probability of receiving the intervention or not. 
It holds by design in randomized controlled trials, while it can be violated in observational studies. 
Assumption \textbf{VC} (VC Class) assumes that the policy class cannot shatter infinitely many points or, more intuitively, it cannot be infinitely complex.
Define $\mathcal{Q}_\mathrm{KT}$ as the state space that satisfies Assumption \ref{ass:kt_18}.

\cite{kitagawa_who_2018} study the properties of EWM. This is defined as:

\begin{equation}\label{eq:ewm}
    \hat{d}_{\mathrm{EWM}} = \argmax_{d \in \mathcal{D}_v} \left\{\hat{W}_n(d) := \frac{1}{n}\sum_{i=1}^n \left[ \frac{Y_i\cdot T_i}{e(X_i)} \cdot d(X_i) + \frac{Y_i \cdot (1-T_i)}{1-e(X_i)} \cdot (1-d(X_i)) \right]\right\}.
\end{equation}

Notice that EWM can be easily rewritten in the notation of Definition \ref{def:policy_choice_problem} by setting $\theta= n$, $\Omega$ as the set of designs that respect Assumption \ref{ass:kt_18}.\textbf{AM}, $m(S^n, d)= \hat{W}_n(d)$, and $\delta(\cdot) = \argmax_{d \in \mathcal{D}_v}\{\cdot\}$. 
If the researcher is collecting data from an experiment she is designing, $\kappa$, the propensity score bound, also enters as a margin of choice: $\theta = (n,\kappa)$.

\begin{proposition}[Regret Bounds \cite{kitagawa_who_2018}]\label{prop:thm_kt18}
Under Assumption \ref{ass:kt_18}, $\hat{d}_{\mathrm{EWM}}$ satisfies:
    \begin{equation}
        \sup_{Q \in \mathcal{Q}_\mathrm{KT}} \mathbb{E}_{Q_{\omega,\theta}}[W_Q(d^*) - W_Q(\hat{d}_{\mathrm{EWM}})] \leq C_1 \frac{M}{\kappa}\sqrt{\frac{v}{n}}.
    \end{equation}
Moreover, for $n \geq 16v$, any $(m, \delta)$ satisfies:
    \begin{equation}
        \sup_{Q \in \mathcal{Q}_\mathrm{KT}} \mathbb{E}_{Q_{\omega,\theta}}[W_Q(d^*) - W_Q(\hat{d})] \geq 2^{-1} \exp\{-2\sqrt{2}\} M \sqrt{\frac{v}{n}}.
    \end{equation}
\end{proposition}

Proposition \ref{prop:thm_kt18} can be summarized into two main points. 
First, the worst-case regret of $\hat{d}_\mathrm{EWM}$ is bounded by (i) a constant that depends on the choice of $n$ and $v$ (and possibly $\kappa$), (ii) state space-dependent constants $M$ and $\kappa$ (respectively, the upper bound on the outcome and on the propensity score), and (iii) a universal constant $C_1$.
Importantly, the rate depends on the ratio between the complexity of the learning problem and the sample size available to solve it. 
Second, the regret upper bound is rate-sharp, as there exists a DGP $Q'\in \mathcal{Q}_\mathrm{KT}$ such that no algorithm could ever achieve a faster rate than $\sqrt{v/n}$.
As a consequence, given $\mathcal{Q}_\mathrm{KT}$, the rate of decay of worst-case regret when choosing $\hat{d}_{\mathrm{EWM}}$ cannot be uniformly improved by any other choice of $(m,\delta)$, keeping fixed $\omega$ and $\theta$ (see Def. \ref{def:policy_choice_problem} for notation).

\subsection{Threads Following Kitagawa and Tetenov (2018)} \label{sec:constrained_ext}
We can group different threads of research that followed EWM depending on the design object taken into consideration.

\subsubsection{Assignment Mechanism}
A first thread of literature has focused on alternative assignment mechanisms and endogenous selection issues.
\paragraph{Policy Learning with Observational Data.}
\cite{athey_policy_2021} asks whether similar regret guarantees can still be obtained when policy learning is based on observational data where propensity scores are unknown, and intervention assignment may be endogenous.
The main contribution is showing that this is possible whenever the welfare of assigning interventions can be estimated through semiparametrically efficient doubly robust scores.
Let $Z_i$ denote the observables used to identify the intervention effect, with $Z_i=T_i$ under selection on observables and $Z_i$ a valid instrument under endogenous assignment.

\begin{assumption}[\cite{athey_policy_2021}'s state space]\label{ass:aw_21} $ $
    \begin{itemize}
        \item[\textbf{ID}.] The intervention is binary $t\in\{0,1\}$. There exist measurable functions $\mu_Q:\mathcal{X}\times\mathcal{T}\to\mathbb{R}$, $\tau_Q:\mathcal{X}\times\mathcal{T}\to\mathbb{R}$, and $g_Q:\mathcal{X}\times \mathrm{supp}(Z_i)\to\mathbb{R}$ such that, defining
        \begin{equation}
            \bar{\tau}_Q(x):=\mathbb{E}_Q[\tau_Q(X_i,T_i)\mid X_i=x]
        \end{equation}
        and
        \begin{equation}
            \Gamma_Q(S_i):=\tau_Q(X_i,T_i)+g_Q(X_i,Z_i)\left[Y_i-\mu_Q(X_i,T_i)\right],
        \end{equation}
        it holds that
        \begin{equation}
            W_Q(d)=\mathbb{E}_Q[d(X_i)\bar{\tau}_Q(X_i)] \quad \forall d\in\mathcal{D}_v
        \end{equation}
        and
        \begin{equation}
            \mathbb{E}_Q[\Gamma_Q(S_i)\mid X_i]=\bar{\tau}_Q(X_i).
        \end{equation}
        \item[\textbf{NE}.] There exist estimators $\hat{\mu}_n$, $\hat{\tau}_n$, and $\hat{g}_n$ such that
        \begin{align}
            & \sup_{x,t}\left|\hat{\mu}_n(x,t)-\mu_Q(x,t)\right| = o_p(1), \\
            & \sup_{x,t}\left|\hat{\tau}_n(x,t)-\tau_Q(x,t)\right| = o_p(1), \\
            & \sup_{x,z}\left|\hat{g}_n(x,z)-g_Q(x,z)\right| = o_p(1),
        \end{align}
        and there exist $\zeta_\mu,\zeta_g \in (0,1)$ and a sequence $a_n\to 0$ such that $\zeta_\mu+\zeta_g\geq 1$ and
        \begin{align}
            \mathbb{E}_Q\left[\left(\hat{\mu}_n(X_i,T_i)-\mu_Q(X_i,T_i)\right)^2\right] & \leq a_n n^{-\zeta_\mu}, \\
            \mathbb{E}_Q\left[\left(\hat{\tau}_n(X_i,T_i)-\tau_Q(X_i,T_i)\right)^2\right] & \leq a_n n^{-\zeta_\mu}, \\
            \mathbb{E}_Q\left[\left(\hat{g}_n(X_i,Z_i)-g_Q(X_i,Z_i)\right)^2\right] & \leq a_n n^{-\zeta_g}.
        \end{align}
        \item[\textbf{VC}.] There exist $\beta\in(0,1/2)$ and $N\geq 1$ such that $v_n := \mathrm{VC}(\mathcal{D}_{v_n}) \leq n^\beta \qquad \forall n\geq N.$
        \item[\textbf{BW}.] There exists $\eta>0$ such that
        \begin{equation}
            |g_Q(x,z)| \leq \eta^{-1} \qquad \forall (x,z)\in \mathcal{X}\times \mathrm{supp}(Z_i).
        \end{equation}
    \end{itemize}
\end{assumption}

Assumption \textbf{ID} (Identification) is the key departure from Assumption \ref{ass:kt_18}: instead of requiring a known randomized assignment rule, the paper defines a state space through semiparametric identification conditions that allow the intervention effect to be recovered from observational data.
Assumption \textbf{NE} (Nuisance Estimation) requires the researcher to estimate first-step objects, such as outcome regressions, propensity scores, or compliance scores, accurately enough for the final score to be asymptotically well behaved.
Assumption \textbf{VC} (VC Class) plays the same role as in \cite{kitagawa_who_2018}, namely controlling the complexity of the policy space.
Finally, Assumption \textbf{BW} (Bounded Weights) generalizes strict overlap: under selection on observables it reduces to bounded inverse propensity weights, while under instrumental variables it rules out weak instruments.
Define as $\mathcal{Q}_\mathrm{AW}$ the set of $Q$s that satisfy Assumption \ref{ass:aw_21}.

Under Assumption \ref{ass:aw_21}, \cite{athey_policy_2021} define a cross-fitted doubly robust welfare estimator:
\begin{equation}\label{eq:aw_dr}
    \hat{W}_n(d) = m_\mathrm{DR}(S^n,d) := \frac{1}{n}\sum_{i=1}^n \left[2d(X_i)-1\right]\hat{\Gamma}_i,
\end{equation}
where $\hat{\Gamma}_i$ denotes a cross-fitted doubly robust score for the conditional gain from treating unit $i$.
Then the corresponding policy rule is:
\begin{equation}
    \hat{d}_\mathrm{DR} = \delta_\mathrm{DR}(m(S^n,d)) := \arg\max_{d \in \mathcal{D}_v}\hat{W}_n(d).
\end{equation}
Notice that, relative to EWM, the choice rule $\delta(\cdot)$ is unchanged: the novelty lies in the construction of $m(S^n,d)$, which now depends on first-step machine-learning estimators of nuisance components.
In the notation of Definition \ref{def:policy_choice_problem}, this corresponds to setting $\theta = n$, letting $\omega$ denote an observational data collection process satisfying Assumption \ref{ass:aw_21}.\textbf{ID}, setting $m(S^n,d)=\hat{W}_n(d)$, and choosing $\delta(\cdot)=\arg\max_{d\in\mathcal{D}_v}\{\cdot\}$.
Define moreover
\begin{equation}
    S_{Q_n}^* := \inf_{d \in \mathcal{D}_{v_n}} \mathrm{Var}_{Q_n}\left(\left[2d(X_i)-1\right]\Gamma_{Q_n}(S_i)\right).
\end{equation}
Under standard regularity conditions, $S_{Q_n}^*$ is the semiparametric efficiency bound for evaluating the best policy in the class.

\begin{proposition}[Asymptotic regret bounds \cite{athey_policy_2021}]\label{prop:aw21}
For any sequence $\{Q_n\}_{n\geq 1}\subseteq \mathcal{Q}_\mathrm{AW}$, $\hat{d}_\mathrm{DR}$ satisfies:
\begin{equation}
    R_{Q_n}(\hat{d}_\mathrm{DR}) = O\left(\sqrt{\frac{v_n S_{Q_n}^*}{n}}\right),
\end{equation}
up to logarithmic terms.
\end{proposition}
Proposition \ref{prop:aw21} shows the same leading dependence on the ratio between policy complexity and sample size shown in \ref{prop:thm_kt18}: regret decays at the $\sqrt{v_n/n}$ rate, up to state-space-dependent constants and logarithmic terms.
\begin{remark}
    All the guarantees in \cite{athey_policy_2021}, including Proposition \ref{prop:aw21}, differ conceptually from the minimax guarantees emphasized in the rest of this review.
    In particular, they do not take the form $\sup_{Q \in \mathcal{Q}}R_Q(\hat d)$ for a fixed state space $\mathcal{Q}$.
    Rather, they are pointwise asymptotic: for each admissible sequence of DGPs $\{Q_n\}_{n\geq 1}$ satisfying Assumption \ref{ass:aw_21}, regret is shown to vanish at the stated rate as $n$ grows.
    The notation $Q_n$ is needed because the paper allows the difficulty of the problem to vary with sample size, through both the complexity of the policy class $v_n$, and the quality of nuisance estimation $S_{Q_n}^*$.
    As a consequence, \cite{athey_policy_2021} fits the general framework of Definition \ref{def:policy_choice_problem}, but it does not deliver a uniform minimax guarantee over a fixed state space in the sense of Eq. \ref{eq:lower_upper}.
\end{remark}
\begin{remark}
    The assignment mechanism allowed by $\mathcal{Q}_\mathrm{AW}$ is richer than in \cite{kitagawa_who_2018}, as it allows for unknown propensity scores, observational designs, and endogenous intervention assignment under valid instruments.
    The price of this generality is that the guarantees are asymptotic and depend on nuisance-estimation rates and on the efficiency bound $S_{Q_n}^*$, rather than only on bounded outcomes and known overlap constants.
    Moreover, all results are conditional on Assumptions \ref{ass:aw_21}.\textbf{ID} and \ref{ass:aw_21}.\textbf{NE}, which impose implicit restrictions on the pairs $(\mathcal{Q},m(S,d))$ that are allowed. 
    In the binary intervention case under selection on observables, the paper also provides a lower bound showing that this rate is sharp up to constants for local asymptotic sequences of problems.
\end{remark}

The results in \cite{athey_policy_2021} were very influential in subsequent literature studying how specific extensions of EWM would behave when propensity scores were unknown.

\paragraph{Policy Learning with Network Interference.}
\cite{viviano_policy_2024} extends empirical welfare maximization to settings with network spillovers. Relative to \cite{kitagawa_who_2018} and \cite{athey_policy_2021}, the main difference is that a unit's outcome may depend not only on her own intervention, but also on the interventions assigned to her neighbours. A policy therefore determines both each unit's intervention and her exposure to treated neighbours. Consequently, welfare depends on the policy profile over the entire network, and observations are no longer independent.

The researcher may observe only a sample of units, together with information about their neighbours. Identification requires sampling to be exogenous conditional on observables, intervention assignment to be conditionally unconfounded, and sufficient overlap for both own intervention status and network exposure. The policy class is assumed to have finite VC dimension. In addition, the network cannot become too dense relative to the effective sample size. Formally, if $N_n$ measures the maximum neighbourhood size, $\delta_n$ is the smallest probability of observing a relevant exposure, and $n_e$ is the expected number of sampled units, the paper requires
\begin{equation}
    \frac{N_n^{3/2}\log(N_n)}{\delta_n}
    =O(n_e^{1/2-\xi})
\end{equation}
for some $\xi\in(0,1/2]$.

\cite{viviano_policy_2024} proposes Network Empirical Welfare Maximization (NEWM), which maximizes an augmented inverse-probability-weighted estimate of welfare. The estimator reweights each sampled unit by the probability of observing the intervention and network exposure induced by a candidate policy, while allowing for a regression adjustment. For many policy classes, the resulting optimization problem can be formulated as a mixed-integer linear program.

With known propensity scores and a bounded regression adjustment, the regret of NEWM is of order
\begin{equation}
    \frac{N_n^{3/2}\sqrt{\log(N_n)v}}
         {\delta_n\sqrt{n_e}},
\end{equation}
where $v$ is the VC dimension of the policy class. Under the network sparsity condition, regret is therefore $O(n_e^{-\xi})$ and, when the maximum degree is uniformly bounded, recovers the usual $1/\sqrt{n_e}$ rate. Thus, network interference affects policy learning through both network density and exposure overlap. The paper also establishes a corresponding lower bound and proposes a network cross-fitting procedure for estimated nuisance components.
\subsubsection{Policy Space}
A second thread of literature has focused on how to choose the policy space $\mathcal{D}_v$ optimally, and on the properties of special cases.

\paragraph{Model Selection for Policy Learning.} \cite{mbakop_model_2021} studies the problem of selecting optimally the complexity of the policy space.
Formally, they consider sieves of assignment rules
\begin{equation}
    \mathcal{D}_1 \subseteq \mathcal{D}_2 \subseteq \mathcal{D}_k \subseteq ... \subseteq \mathcal{D}_K
\end{equation}
of increasing VC dimension.
One key insight is that, given this sieve characterization, we can decompose regret as
\begin{align}
    R_Q(\hat{d}) & = \mathbb{E}_{Q_{\omega,\theta}}[W_Q(d^*) - W_Q(\hat{d}_k)] \\
    & = \mathbb{E}_{Q_{\omega,\theta}}[W_Q(d^*) - W_Q(d^*_k)] + \mathbb{E}_{Q_{\omega,\theta}}[W_Q(d^*_k) - W_Q(\hat{d}_k)] \\
    & = \underbrace{W_Q(d^*) - W_Q(d^*_k)}_{\text{approximation error}} + \underbrace{\mathbb{E}_{Q_{\omega,\theta}}[W_Q(d^*_k) - W_Q(\hat{d}_k)]}_{\text{estimation error}}.
\end{align}
Here, $d^*$ denotes the oracle policy in the collection and $d^*_k$ the oracle policy in the sieve at level $k$.
This decomposition highlights an estimation-approximation error trade-off.
On the one hand, the more complex the policy space is, the closer the relative oracle $d^*_k$ will be to the unconstrained optimum $d^*$, and therefore the lower the approximation error.
On the other hand, higher complexity leads to higher estimation error both pointwise and in the worst case.
To trade these two errors efficiently, \cite{mbakop_model_2021} define $\hat{d}_\mathrm{PWM}$ as:
\begin{equation}
    \hat{d}_\mathrm{PWM}:= \arg \max_{1 \leq k \leq K} \arg \max_{d \in \mathcal{D}_k} \left\{\hat{W}_n(d) - C_n(k) - \sqrt{\frac{t_k}{n}} \right\},
\end{equation} 
where $C_n(k)$ denotes a cost function that penalizes the policy space complexity, and $\sqrt{t_k/n}$ is a technical device used to ensure that, if $K$ is not finite, the classes get penalized at a sufficiently fast rate as $k$ increases. 
\begin{example}
One convenient example for $C_n(k)$ is a holdout penalty. 
In particular, let
\begin{equation}
    C_q(k) := \hat{W}_q(\hat{d}_{\mathrm{EWM},k}) - \hat{W}_r(\hat{d}_{\mathrm{EWM},k})
\end{equation}
where $q$ and $r$ denote two mutually exclusive subsamples $S_q$ (training sample) and $S_r$ (test sample) of $S^n$. 
Let $\hat{d}_{\mathrm{EWM},k}$ be the EWM policy for sieve $\mathcal{D}_k$ estimated in the subsample $S_q$.
Then, we can rewrite the PWM objective as:
\begin{equation}
    \hat{d}_\mathrm{PWM} =\arg \max_{1 \leq k \leq K}  \left\{\hat{W}_r(\hat{d}_{\mathrm{EWM},k}) - \sqrt{\frac{t_k}{n}} \right\},
\end{equation}
that is the policy space that maximizes out-of-sample empirical welfare.
\end{example}
Let's now map this framework back onto Definition \ref{def:policy_choice_problem}. 
In particular, PWM is defined by the estimator 
\begin{equation}
m_\mathrm{PWM}(S^n, d) = \hat{W}_n(d) - C_n(k) - \sqrt{\frac{t_k}{n}}
\end{equation} 
and the choice rule 
\begin{equation}
    \delta_\mathrm{PWM}(\cdot) = \arg \max_{k \in \mathcal{K}, d \in \mathcal{D}_k}\{\cdot\}. 
\end{equation}
The margin of choice for data collection is $\theta = n$ and $\omega$ is required to respect unconfoundedness.
The state space is the one considered in \cite{kitagawa_who_2018}\footnote{See Assumption \ref{ass:kt_18} for the formal definition.}, with additional structure constraining the complexity penalizer $C_n(k)$ and the sieves.
Formally, $\mathcal{Q}_\mathrm{MT}$, compared to $\mathcal{Q}_\mathrm{KT}$, further imposes that there exist positive constants $c_0$ and $c_1$ such that $C_n(k)$ satisfies the following tail inequality for every $n, k$, and for every $\epsilon>0$
\begin{equation}
    \sup_{Q \in \mathcal{Q}_\mathrm{MT}} \mathbb{P}_{Q_{\omega,\theta}}\big[\hat{W}_n(\hat{d}_{\mathrm{EWM},k}) - W(\hat{d}_{\mathrm{EWM},k}) - C_n(k)>\varepsilon\big] \leq c_1 \exp \{-2c_0n\epsilon^2\},
\end{equation}
and that there exists a universal constant $C_1$ such that, for every $n$, $C_n(k)$ satisfies
\begin{equation}
    \sup_{Q \in \mathcal{Q}_\mathrm{MT}} \mathbb{E}_{Q_{\omega,\theta}}[C_n(k)]\leq C_1 \sqrt{\frac{v_k}{n}}.
\end{equation}
\cite{mbakop_model_2021} show that, provided that $d^* \in \mathcal{D}$, for a finite $K$, 
\begin{equation} 
    \sup_{Q \in \mathcal{Q}_\mathrm{MT}} \mathbb{E}_{Q_{\omega,\theta}}[W_Q(d^*) - W_Q(\hat{d}_\mathrm{PWM})] \leq C_1 \sqrt{\frac{v_k}{n}} + \sqrt{\frac{\log(c_1 K e)}{n}}
\end{equation}
where $v_k$ denotes the VC dimension of $\mathcal{D}_k$ that contains $d^*$.
This is a powerful result as it shows that, under the additional conditions stated above, PWM achieves the minimax rate proved in \cite{kitagawa_who_2018}. 
As a result, a policymaker who does not have a preference for a specific policy space can let the data identify the best among competing spaces, optimally trading off approximation and estimation error.
The optimality is implied by the fact that PWM achieves the minimax rate, which no alternative algorithm can uniformly improve.

\paragraph{Extensions of PEWM.} Recent papers extend the rationale of \cite{mbakop_model_2021} to richer observational settings within the doubly robust framework of \cite{athey_policy_2021}. 
\cite{fang_2025} extends this logic to observational settings with multivalued interventions and unknown propensity scores. 
Relative to \cite{mbakop_model_2021}, the main innovation is replacing the empirical welfare criterion with a cross-fitted doubly robust welfare estimator \citep[as in][]{athey_policy_2021}, and then selecting among sieve policy classes using either Rademacher or holdout penalties. 
They derive oracle inequalities showing that the resulting data-driven rule trades off approximation and estimation error as if these quantities were known, and illustrate how to implement the sieve selection with monotone single-index rules and discretizations of smooth policy functions based on linear sieves or deep neural networks.

\cite{ai_2026} extends this logic to continuous interventions. 
In that setting, welfare can no longer be evaluated by a simple sample average, so $m(S^n,d)$ must approximate the value of each continuous level by smoothing across nearby intervention levels. 
The main contribution is therefore to show how the PEWM can be extended when one must jointly choose policy complexity and a bandwidth parameter. 
The paper develops a penalized procedure that automates both choices using the data. 
When propensity scores are unknown, a double-debiased version \citep{athey_policy_2021} yields a comparable guarantee, so the paper can be read as the continuous-intervention counterpart to the multivalued-intervention extension in \cite{fang_2025}.

\cite{Ponomarev2026} studies a closely related adaptive choice of policy complexity in observational settings, while keeping focus on utilitarian welfare and binary choice rules. 
Compared to \cite{mbakop_model_2021}, their contribution is mainly theoretical: combining doubly robust welfare estimation with sample splitting, they derive sharper finite-sample upper bounds on expected regret, a matching lower bound up to constants, and therefore show that Adaptive Welfare Maximization is minimax-rate optimal while achieving nearly-oracle performance in selecting the relevant policy class.

\paragraph{Policy Learning with Unobserved Heterogeneity.} \cite{opocher_2026} studies the case in which intervention effects vary with a latent characteristic $A_i$ that is not directly observed, but only through a noisy proxy $\hat{A}_i$. 
Relative to \cite{kitagawa_who_2018}, the assignment mechanism and utilitarian welfare are unchanged, but the policy space can either ignore this source of heterogeneity and use rules $d(X_i)$, or augment targeting with the proxy and use rules $d(X_i,\hat{A}_i)$. 
The paper evaluates both classes relative to an oracle that observes $A_i$ and derives rate-sharp regret bounds for each. 
For covariate-based rules, worst-case regret equals the usual statistical term plus an approximation term reflecting the residual intervention-effect heterogeneity left unexplained by $X_i$. 
For proxy-augmented rules, worst-case regret equals the usual statistical term plus a noise term proportional to the root mean-squared error of $\hat{A}_i$. 
Therefore, richer targeting variables improve policy learning only if the welfare gain from capturing latent heterogeneity outweighs both the increase in policy-class complexity and the noise introduced by measuring the proxy.
In the notation of Def. \ref{def:policy_choice_problem}, this means that the collection parameter space $\Theta_\omega$ collects the sample size of the experiment (which defines $\omega$) and the \textit{quality of information} $t$ that the researcher has about $A_i$. 
Examples of $t$ include the number of measurements available in a repeated-measurement setting, the sample size used to learn $\hat{A}_i$ when it is estimated with auxiliary data, or the image quality when $\hat{A}_i$ is measured with satellite images.
The paper then uses these bounds to study the data-collection problem over $\Theta_\omega = (n,t)$: under a budget constraint that specifies the cost of $n$ and $t$, and the available budget, how should the policymaker trade off improving the precision of $\hat{A}_i$ and increasing the policy-learning sample size? 
It derives sufficient conditions under which either the covariate-based design or an augmented design with information level $t$ is minimax-dominant in the sense of Eq. \ref{eq:minimax_dom}.

\paragraph{Choosing Who Chooses.} \cite{kitagawa_choosing_who_choses} expands the policy space in a different direction. 
Relative to the binary assignment problem in \cite{kitagawa_who_2018}, the planner now chooses among three policies for each covariate value: assigning the intervention, assigning the status quo, or letting the unit self-select into intervention. 
The paper compares \textit{paternalistic} (top-down assignment) and \textit{laissez-faire} (bottom-up self-selection) approaches to policymaking.
Its main conceptual point is that mixing the two approaches across the population (i.e. deciding separately for each covariate value which approach works best) outperforms both approaches when considered separately. 
The optimal policy is estimated via empirical welfare maximization, implemented over policy trees, on this richer action space. 
The paper further uses LATEs for takers and non-takers to interpret why self-selection raises welfare in some values of the covariate space and lowers it in others.

\paragraph{Dynamic Assignment.} \cite{Shosei_2025} extends EWM to allow for dynamic policy choices that map units' histories into intervention assignments over multiple stages. 
Relative to \cite{kitagawa_who_2018}, the data collection process $\omega$ is required to satisfy sequential unconfoundedness, while the rest of the state space preserves the same boundedness, overlap, and complexity assumptions (see Assumption \ref{ass:kt_18}). 
The paper proposes two dynamic EWM procedures, one based on backward induction and one based on simultaneous maximization over the whole regime. 
The first is computationally lighter, but is guaranteed to recover the optimal constrained regime only when the first-best continuation rules belong to the feasible class at later stages; the second is computationally harder, but remains consistent for the best regime within the constrained class without that requirement. 
For both procedures the paper derives finite-sample, worst-case regret bounds and shows that, in the experimental setting, worst-case regret decays at the minimax-optimal rate of \citet{kitagawa_who_2018}. 
It further shows how the simultaneous procedure can accommodate intertemporal budget or capacity constraints.

\paragraph{Matching Policies.} \cite{hazard_kitagawa_2025} studies a different policy object: rather than assigning one intervention to each unit, the planner matches units on one side of the market to units on the other side. 
Relative to \cite{kitagawa_who_2018}, the policy is therefore not a rule $d:\mathcal{X}\to\mathcal{T}$ but a feasible matching policy $\pi_n$ over pairs $(X_i,X_j)$ satisfying one-to-one constraints. 
The paper estimates an average match cost $c(X_i,X_j)$ from training data and then chooses $\hat{\pi}_n$ by entropy-regularized empirical optimal transport. 
Its main result is a non-asymptotic upper bound on regret relative to the oracle unregularized matching policy. 
Under bounded costs, expected regret is bounded by the estimation error of $\hat{c}$ plus a regularization-bias term of order $\log n/\eta$, where $n$ is the size of the matching market and $\eta$ indexes the strength of regularization. 
The paper emphasises the computational tractability of the method but does not provide a matching lower bound on regret.

\subsubsection{Welfare Function} 
A third thread of literature has considered different welfare objectives that may interest a real-world policymaker.

\paragraph{Equality-Minded EWM.} \cite{kitagawa_equality-minded_2021} extended EWM to allow for an equality-minded objective function. 
In particular, they consider a welfare function that satisfies the Pigou-Dalton principle of transfers: a transfer of (potential) outcome from a higher-ranked individual to a lower-ranked individual is always desirable when it does not change their ranks in the status quo.
While satisfying this constraint, a policymaker is allowed to \textit{prioritize} (meaning place more weight on) lower ranks of the outcome's distribution.

Formally, they redefine welfare as:
\begin{equation}
    W_Q^\mathrm{eq}(d) = \int_0^\infty \Delta\left(F(y,d)\right) \mathrm{d}y,
\end{equation}
where $F(y,d)$ is the quantile function of the outcome under policy $d$ and $\Delta: [0,1] \to [0,1]$ is a nonincreasing, nonnegative function with $\Delta(0)= 1$ and $\Delta(1) = 0$.
$W_Q^\mathrm{eq}(d)$ satisfies the Pigou-Dalton principle if and only if $\Delta(\cdot)$ is convex.
Regret is then defined as
\begin{equation}
    R_Q(d) = \mathbb{E}_Q\left[\max_{d \in \mathcal{D}_v} W_Q^\mathrm{eq}(d) - W_Q^\mathrm{eq}(\hat{d})\right].
\end{equation}
They define a new welfare estimator:
\begin{equation}
    m_\mathrm{EM}(S, d) = \hat{W}(d) :=\int_0^\infty \Delta \left(\max\{\hat{F}(y,d), 0\} \right) dy, 
\end{equation}
where
\begin{equation}
    \hat{F}(y,d) := 1 - \frac{1}{n} \sum_{i = 1}^n \left[ \frac{T_i}{e(X_i)} \cdot d(X_i) + \frac{1-T_i}{1-e(X_i)} (1-d(X_i)) \right]\cdot \mathbf{1}\{Y_i \geq y\}.
\end{equation}
Finally, they define Equality-Minded EWM (EM-EWM) as:
\begin{equation}
    \hat{d}_\mathrm{EM} = \delta_\mathrm{EM}(m(S,d)) := \arg \max_{d \in \mathcal{D}} \hat{W}(d).
\end{equation}
The main result of the paper mimics the main result in \cite{kitagawa_who_2018}. 
They show that, under the assumptions that (i) $\Delta$ is nonincreasing and convex, with a finite first derivative at the origin; (ii) the policy class $\mathcal{D}_v$ has finite VC dimension; and (iii) strict overlap, unconfoundedness, and concentration of the outcome hold, $(m_\mathrm{EM}, \delta_\mathrm{EM})$ attains the minimax rate.
Interestingly, the minimax rate is the same as the utilitarian welfare case, $\sqrt{v/n}$, as all the welfare losses that come from being equality-minded decay exponentially with $n$.

Two recent papers push this logic further. \cite{fan_qi_xu_2025} keeps the group-agnostic flavor of EM-EWM but sharpens its distributional concern by replacing the rank-dependent welfare functional with an $\alpha$-expected welfare criterion that maximizes the average welfare of the worst-off $\alpha$-fraction of the population, thereby interpolating between utilitarian EWM and a Rawlsian objective; they propose a debiased estimator and derive asymptotic regret bounds and inference for optimal welfare. \cite{terschuur_2025} instead abstracts from the rank-dependent class used in EM-EWM and studies policy learning with general semiparametric social welfare functions estimable by locally robust orthogonal moments and U-statistics, a framework that accommodates inequality-aware, inequality-of-opportunity-aware, and intergenerational-mobility objectives while still delivering asymptotic regret guarantees.

\paragraph{Fair Policy Learning.} 
\cite{viviano_fair_2024} extends EWM to policy makers who are willing to sacrifice some aggregate welfare to protect groups defined by a sensitive attribute $S_i$. Rather than maximizing a fixed weighted average of group welfare, the policy maker first restricts attention to the Pareto frontier: policies for which one group's welfare cannot be improved without reducing that of another group. With two groups, this frontier can be traced by maximizing
\begin{equation}    
    \alpha W_Q(d,S_i=1)+(1-\alpha)W_Q(d,S_i=0),
\end{equation}
for different values of $\alpha\in(0,1)$. Among the resulting Pareto-efficient policies, the policy maker selects the one that minimizes a chosen measure of unfairness, denoted by $U_Q(d)$.

The framework accommodates several notions of unfairness. These include differences across groups in the probability of receiving the intervention, disparities in the welfare generated by the policy, and violations of incentive compatibility measured by the benefits that individuals could obtain by misreporting their sensitive attribute.

Because the Pareto frontier is unknown, it must be estimated from the data. The authors propose an algorithm that approximates this frontier and selects the estimated policy with the lowest unfairness. Under unconfoundedness, strict overlap, bounded outcomes, a policy class with finite VC dimension, and sufficiently accurate estimation of the conditional means and propensity scores, the resulting unfairness regret converges at rate $1/\sqrt{n}$. A corresponding lower bound shows that no data-dependent procedure can achieve a uniformly faster rate. The proposed policy rule is therefore minimax-rate optimal.
\paragraph{Policy Learning with Random Constraints.} 
\cite{sun_2026} considers the case where the budget constraint of Definition \ref{def:policy_choice_problem} is uncertain and therefore needs to be estimated.

Formally, the cost of collecting data is not considered (in the notation of Definition \ref{def:policy_choice_problem}, $c(\omega,\theta,Q)=0$), and the cost of implementing a policy $d$ is random: $\sigma(d,Q)=\sigma_i$.
Then, the decision problem can be re-written as:
\begin{align}
    & \max_{d \in \mathcal{D}_v} W_Q(d) \\
    & \text{s.to: } \mathbb{E}_Q[\sigma_i\cdot d(X_i)] \leq B_0
\end{align}
Define $\Sigma_Q(d) := \mathbb{E}_Q[\sigma_i\cdot d(X_i)]$.
\cite{sun_2026} focuses on uniform asymptotic efficiency:
\begin{equation}
    \lim_{n\to \infty} \sup_{Q \in \mathcal{Q}} \mathbb{P}_{Q_{\omega,\theta}}\left\{ W_Q(\hat{d}) - W_Q(d^*) < -\epsilon \right\} = 0,
\end{equation} 
for any $\epsilon > 0$, and uniform asymptotic feasibility:
\begin{equation}
    \lim_{n\to \infty} \sup_{Q \in \mathcal{Q}} \mathbb{P}_{Q_{\omega,\theta}}\left\{ \Sigma_Q(\hat{d}) > B_0 \right\} = 0.
\end{equation} 
The first important result in \cite{sun_2026} is that, for a sufficiently rich state space (such as $\mathcal{Q}_\mathrm{KT}$), it is impossible for any $(m,\delta)$ to achieve both uniform asymptotic efficiency and feasibility.
A second result is that the trivial extension of EWM that just replaces $\Sigma_Q(d)$ with the sample analogue is neither uniformly asymptotically welfare-efficient nor uniformly asymptotically feasible.
Finally, \cite{sun_2026} introduces the new welfare function $W_Q^\mathrm{TO}$ that trades off units of welfare for budget overrun:
\begin{equation}
    W_Q^\mathrm{TO} := W_Q(d) - \frac{\bar{\lambda}}{r} \cdot (\Sigma_Q(d) - B_0)_+
\end{equation}
The intuition is that the PM can decide to borrow the quantity $(\Sigma_Q(d) - B_0)_+$ at a cost $\bar{\lambda}$, which can be measured as a loss in units of welfare with the scaling factor $r$.
Then, the paper defines a trade-off rule as:
\begin{equation}
    \hat{d}_\mathrm{TO} := \arg\max_{d \in \mathcal{D}} \left\{ \hat{W}_n(d) - \frac{\bar{\lambda}}{r} \cdot (\hat{\Sigma}_n(d) - B_0)_+ \right\}
\end{equation}
The trade-off rule is shown to achieve uniform asymptotic efficiency and to guarantee a finite bound on the budget overrun.
However, $\hat{d}_\mathrm{TO}$ is not proven to be minimax optimal or minimax-rate optimal.

\section{Interesting Avenues}\label{sec:interesting_venues}
In this section, I highlight some areas at the frontier of the literature that I personally find interesting.

\subsection{Evidence Aggregation}
\cite{ishihara_evidence_2026} study policy choice when the planner has no direct sample from the target population, but only a collection of external studies reporting effect estimates $\hat\tau_k\sim\mathcal{N}(\tau_k,\sigma_k^2)$ for related populations, $k=1,\dots,K$. 
Writing $\tau_0$ for the target-population welfare effect, they restrict attention to non-randomized linear aggregation rules
\begin{equation}
    \hat d_{\mathrm{IK}}
    :=
    \delta_{\mathrm{IK}}\{D\}
    =
    \mathbf{1}\left\{\sum_{k=1}^K w_k\hat\tau_k\geq 0\right\},
    \qquad
    \sum_{k=1}^K w_k=1,
\end{equation}
and assume that the feasible set for $(\tau_0,\tau_1,\dots,\tau_K)$ is symmetric and invariant to common shifts. 
Under these conditions, maximum regret depends on the weights only through the worst-case bias
\begin{equation}
    b(w):=
    \max_{\tau\in\mathcal{T}:\tau_0=0}
    \sum_{k=1}^K w_k(\tau_k-\tau_0)
\end{equation}
and the sampling standard deviation
\begin{equation}
    s(w):=\left(\sum_{k=1}^K w_k^2\sigma_k^2\right)^{1/2}.
\end{equation}
Their main theorem shows that the minimax-regret aggregation weights solve
\begin{equation}
    w_{\mathrm{IK}}^*
    \in
    \arg\min_w
    s(w)\eta\!\left(\frac{b(w)}{s(w)}\right),
    \qquad
    \eta(a):=\max_{t\geq 0} t\Phi(-t+a).
\end{equation}
The key point is that evidence aggregation becomes a bias-variance problem tailored to regret rather than estimation error: relative to the minimax-MSE rule, minimax regret places more emphasis on controlling worst-case bias in extrapolating from study populations to the target one. 
For meta-regression and Lipschitz-type parameter spaces, the bias term $b(w)$ can be computed by linear programming, which makes the rule operational in applications.

\paragraph{Follow-up Questions.}
As the authors acknowledge, one follow-up question arising from this study is how to account for publication bias.
I divide publication bias into two layers.
First, I consider the editor as an agent who assigns different publication probabilities depending on the statistical significance of results, e.g. whether $\hat{\tau}_k/\sigma_k > 1.96$.
This would make the actual distribution of observed evidence a truncated normal and would keep the problem tractable. 
Second, I consider the researcher as an agent who can design protocols (e.g. data-cleaning algorithms) such that $\hat{\tau}_k/\sigma_k > 1.96$ by design.
This would maximize the probability of publishing the paper by backward induction.
In practice this could happen in many ways. 
For example, if the algorithm chosen by the researcher to estimate $\tau$ is sufficiently unstable \citep[in the sense of][]{bousquet}, they could trim the data until the result becomes statistically significant. 

This scenario opens several questions.
How can optimal evidence-aggregation weights account for this distortion? 
Should the policymaker ignore some studies if there is evidence of manipulation?
Should the policymaker set the weights ex-ante and also be minimax against manipulation, or should they instead let the weights \textit{adapt} to manipulation in the spirit of \citet{armstrong_kline_sun_25}?
At what cost would it be optimal for the policymaker to request the full data from the researcher and independently reassess the results?

\subsection{Implementation}
It is often tempting to think about complex interventions as binary \textit{treatments}.
This is because such a simple framework allows us to provide solid theoretical guarantees and intuitive guidelines for applied researchers.
As an example, consider an information campaign.
We describe as \textit{treated} an individual who knows the information sent in the campaign and as \textit{untreated} an individual who does not know it.
Then, the implementation is the exact way such information is conveyed to individuals.
Examples include a flyer in the mailbox, a text message on WhatsApp, or a video on YouTube.
Simple ways to think of implementation include considering it as an independent dimension, that is, the \textit{way} we intervene on the status quo while keeping the intervention fixed, or as a feature of the intervention itself. 
A more sophisticated way would be to consider the map we adopt to transport an individual from the distribution of their potential outcome under the status quo to the distribution under the intervention.
Either way, the policy recommendation a researcher may produce about the information campaign is informative only for a policymaker who plans to intervene in the full population in exactly the same way.
In other words, implementation is fixed across the learning and implementation stages.
However, it is often not possible for the policymaker to replicate the implementation the researcher adopted in the learning sample in the full population.
This could be the case because of implementation costs or ethical concerns.

This scenario opens several questions.
Should this mismatch between how a researcher and a policymaker can intervene in the status quo be taken into account at the research design stage?
What's the normative value of studies that cannot be implemented? 
How can those contribute to policy choice?
Consider the case in which a policymaker has an implementation in mind and can aggregate evidence across studies that have different implementations as in \cite{ishihara_evidence_2026}. 
Under what conditions is it optimal to aggregate existing evidence while keeping the original target implementation? 
When is it optimal to change the target implementation to be closer to those already tested? 
And when should a policymaker collect their own data and directly test the target implementation while ignoring existing evidence?

\section{How to Produce Recommendations}\label{sec:how_to_produce}
This section is tailored to a more applied audience and has the objective of guiding the steps an applied researcher should take to produce and communicate a policy recommendation.

In the first part, I provide two diagrams that position each theoretical paper in relation to concrete questions and concerns that an applied researcher may consider when solving the research design problem.
This part takes the \textit{ex-ante} perspective described in the rest of the review and can be considered as a map that practitioners can use to navigate the theoretical literature.

In the second part, I consider the \textit{ex-post} perspective of a researcher who has solved the research design problem and collected the data, and now wants to communicate the value of their recommendation in a simple and intuitive way.
I introduce a beta version of \texttt{policytargetr}, a new \texttt{R} package that takes the data and research design as inputs and produces \textit{one table} and \textit{two graphs} that the researcher can plug into the `policy implications' section of their paper.

I present the setting of \citet{hussam_2022} as a working example to show a use case for the diagrams in Section \ref{sec:before_collecting} and the package in Section \ref{sec:after_collecting}.

\subsection{\textit{Before} Collecting Data}\label{sec:before_collecting}

At this stage, the researcher is writing what one may call a \textit{normative} pre-analysis plan. 
This would fix a choice for $(m,\delta)$ before the data collection takes place, following the guidance that the theory provides on the interplay with $(\omega,\theta)$.\footnote{See Definition \ref{def:policy_choice_problem} for reference on notation.}
The data-collection process $\omega$ and its design margins $\theta$ are not fully fixed yet, but the researcher usually has a tentative environment in mind dictated by practical feasibility constraints and wants to understand which choice of $(m,\delta)$ best fits their setting.
Therefore, conditional on this tentative $(\omega,\theta)$, the ex-ante question is which estimator $(m)$ and decision rule $(\delta)$ the theory suggests.

A \textit{normative} pre-analysis plan for policy choice should make explicit at least the following objects:
\begin{enumerate}
    \item The \textbf{policy space}: what set of policies we are considering. \\
    Examples include targeting based on covariates, deciding on full adoption, and considering a specific policy function (e.g. threshold rules or trees).
    \item The \textbf{welfare criterion}: what criterion we would maximize if we were an oracle. \\
    Examples include average welfare, inequality-averse welfare, fairness-aware welfare.
    \item The \textbf{feasible collection processes}: what types of data we can collect. \\
    Examples include experimental data, observational data, data from a distinct population, evidence across different studies.
    \item The \textbf{potential failures}: the set of adversarial conditions the researcher wants the recommendation to be robust against. \\
    Examples include partial internal validity, lack of external validity, sampling uncertainty. 
\end{enumerate}

To mimic the structure of the theoretical review, Figures \ref{fig:unconstrained_tree} and \ref{fig:constrained_tree} help navigate the literature, starting with the conditions that may interest an applied researcher across the four dimensions above.
Figure \ref{fig:unconstrained_tree} refers to an unconstrained policy space (details in Section \ref{sec:unconstrained}). 
Figure \ref{fig:constrained_tree} refers to a constrained policy space (details in Section \ref{sec:constrained}).

A researcher can start from the most binding feature of the environment she expects to face in terms of $(\omega, \Theta_\omega)$ (e.g. the assignment is not random, there may be spillovers, there is partial internal validity) and its interplay with $m$, and then ask what the theory suggests to do when choosing $(m,\delta)$. 
Each leaf then points to the branch of the theory that is most informative for choosing $(m,\delta)$ for that specific concern.

As a working example, consider the setting in \citet{hussam_2022}.
The authors study the effect of providing a cash grant to micro-entrepreneurs on their profits with a randomized controlled trial in rural India. 
This experiment is motivated by the normative objective of fostering economic development in developing countries by investing in local entrepreneurs.
The trial was conducted in the city of Amravati, India, between 2016 and 2018. 
The sample consists of 1,345 micro-entrepreneurs operating informal businesses in retail and services. 
First, participants were assigned to peer groups of five based on geographic proximity. 
Within these groups, individuals were asked to rank their peers on future business outcomes, including future profits and marginal returns to capital. 
Then, one-third of the sample was randomly assigned to receive an unconditional cash grant of 6,000 INR (roughly \$100).
The authors study the effect of the cash grant on various outcomes.
Here we focus on entrepreneurs' profits.

I now highlight three challenges for research design for policy choice that could arise in this setting.\footnote{Unfortunately, it is not possible to confirm that these were actual concerns for the authors, since the pre-analysis plan is not available.}

First, spillovers may be a first-order concern in this environment. 
Entrepreneurs are grouped by geographic proximity and operate in local retail and service markets, so one unit's grant may affect the profits of other units through competition, demand spillovers, imitation, or informal insurance within the peer group. 
If this is a serious ex-ante concern, the relevant branch in Figure \ref{fig:constrained_tree} is the assignment leaf on network spillovers, which points to \citet{viviano_policy_2024}. 
In practice, this means that before collecting data the researcher should decide how to measure the relevant network, what variation in neighbours' intervention exposure the experimental design must generate, and then choose an estimator $m$ and decision rule $\delta$ that are robust to interference rather than only to i.i.d. sampling uncertainty, as suggested in \citet{viviano_policy_2024}.

Second, the PM may care about a target population different from the one observed in Amravati. 
A lender or ministry may be interested in applying the same grant program in another district, another Indian state, or even another country. 
If this is a first-order concern for the researcher, Figure \ref{fig:unconstrained_tree} points to \citet{adjaho_external_2022} and \citet{kido_distributionally_2022}. 
In such cases, the researcher may want their recommendations to be robust over a set of plausible target populations. 
As a result, they may follow \citet{adjaho_external_2022} and \citet{kido_distributionally_2022} to define the welfare function and the empirical analogue to maximize accordingly.
Moreover, this choice affects what covariates and contextual information should be collected to quantify the distance between populations.

Third, the PM may care not only about average profits, but also about how those gains are distributed across the target population. 
A targeting rule based on community rankings could maximize utilitarian welfare and still be unattractive if it systematically disadvantages poorer entrepreneurs, younger entrepreneurs, or protected groups. 
Figure \ref{fig:constrained_tree} highlights two possibilities. 
If the relevant normative concern is inequality in the distribution of outcomes, the appropriate references are \citet{kitagawa_equality-minded_2021}, \citet{fan_qi_xu_2025}, and \citet{terschuur_2025}; these papers would guide the researcher in defining equality-minded welfare objectives and appropriate empirical analogues.
If instead the PM wants explicit fairness constraints across protected groups, \citet{viviano_fair_2024} suggests maximizing among Pareto-optimal allocations over protected groups.

\begin{landscape}
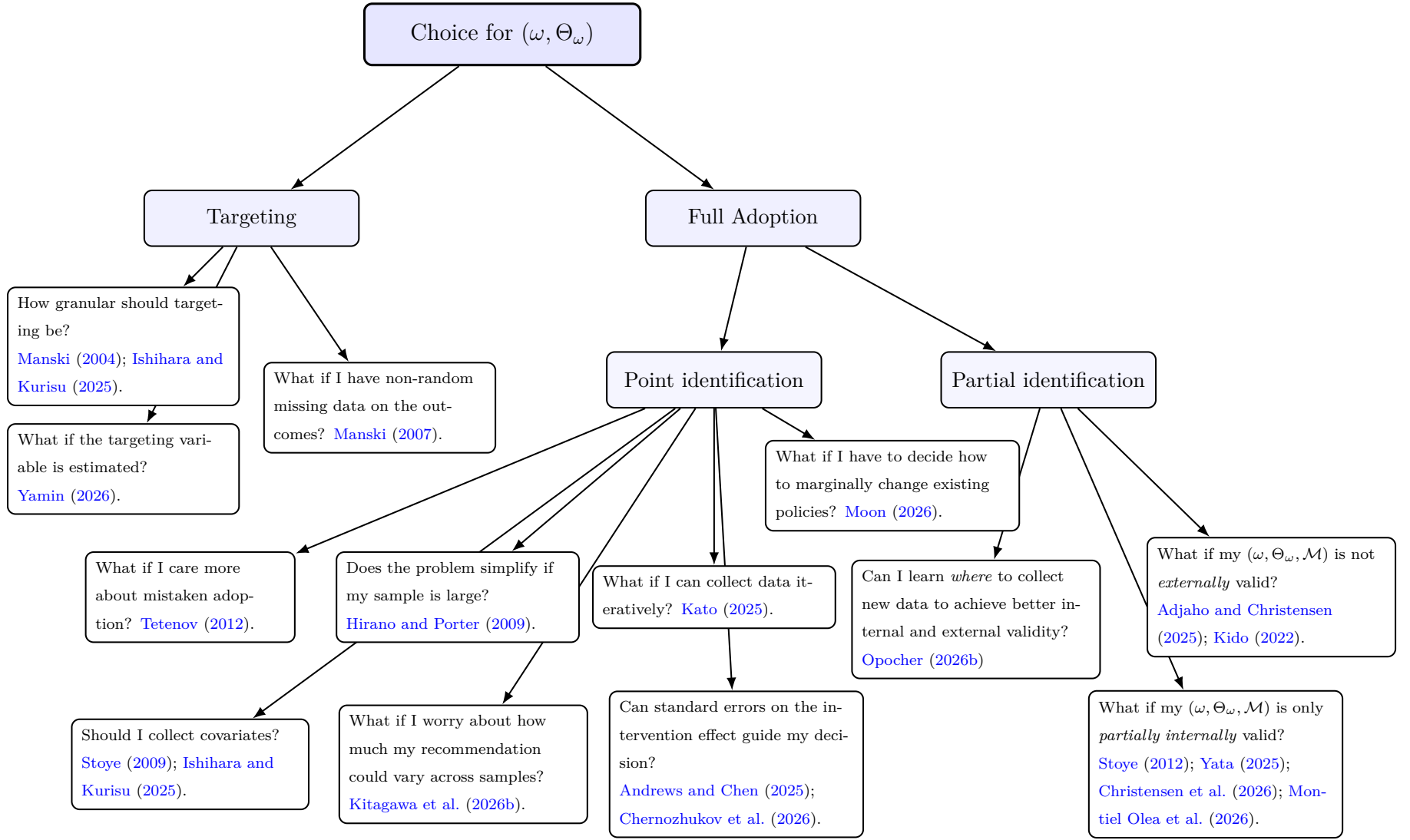
\begin{figure}[p]
\centering
\caption{Designing Research for Unconstrained Policy Choice}
\label{fig:unconstrained_tree}
\scalebox{0.96}{%
\begin{tikzpicture}[
    x=1cm,
    y=1cm,
    arr/.style={-Latex, thick},
    root/.style={draw, rounded corners, very thick, align=center, fill=gray!15, text width=4.6cm, minimum height=1.1cm, inner sep=5pt, font=\normalsize},
    branch/.style={draw, rounded corners, thick, align=center, fill=blue!6, text width=3.5cm, minimum height=1cm, inner sep=5pt, font=\small},
    box/.style={draw, rounded corners, thick, align=left, fill=white, text width=#1, inner sep=5pt, font=\scriptsize},
    box/.default=3.5cm
]
    \node[root, fill=blue!10] (root) at (14.0,15.1) {Choice for $(\omega, \Theta_\omega)$};

    \node[branch, fill=blue!6] (targeting) at (9.5,11.8) {Targeting};
    \node[branch, fill=blue!6] (adoption) at (18.5,11.8) {Full Adoption};
    \draw[arr] (root) -- (targeting);
    \draw[arr] (root) -- (adoption);

    \node[box=3.8cm] (shrinkage_many) at (7.2,7.3) {What if the targeting variable is estimated?\newline \citet{yamin2026}.};
    \draw[arr] (targeting) -- (shrinkage_many);
    \node[box=3.8cm] (targetsmall) at (7.2,9.5) {How granular should targeting be?\newline \citet{manski_statistical_2004}; \citet{Ishihara_2025}.};
    \node[box=3.8cm] (targetmissing) at (11.8,8.4) {What if I have non-random missing data on the outcomes? \citet{MANSKI2007105}.};
    \draw[arr] (targeting) -- (targetsmall);
    \draw[arr] (targeting) -- (targetmissing);

    \node[branch, fill=blue!4] (adoptpoint) at (17.8,8.9) {Point identification};
    \node[branch, fill=blue!4] (adoptpartial) at (23.8,8.9) {Partial identification};
    \draw[arr] (adoption) -- (adoptpoint);
    \draw[arr] (adoption) -- (adoptpartial);

    \node[box=4.2cm] (confidence) at (18.2,2.0) {Can standard errors on the intervention effect guide my decision?\newline \citet{andrews_chen_2025}; \citet{chernozhukov_lee_rosen_sun_2026}.};
    \draw[arr] (adoptpoint) -- (confidence);
    \node[box=3.4cm] (adoptasym) at (8.4,5.0) {What if I care more about mistaken adoption? \citet{tetenov_asymmetric_2012}.};
    \node[box=3.9cm] (covariates_full) at (8.4,2.0) {Should I collect covariates?\newline \citet{stoye_minimax_2009}; \citet{Ishihara_2025}.};
    \draw[arr] (adoptpoint) -- (covariates_full);
    \draw[arr] (adoptpoint) -- (adoptasym);
    \node[box=4.1cm] (variance) at (13.3,2.0) {What if I worry about how much my recommendation could vary across samples?\newline \citet{kitagawa_26_bio}.};
    \draw[arr] (adoptpoint) -- (variance);
    \node[box=4cm] (adoptlocal) at (13.2,5.0) {Does the problem simplify if my sample is large?\newline \citet{hirano_asymptotics_2009}.};
    \node[box=4.0cm] (adoptadaptive) at (17.8,5.0) {What if I can collect data iteratively? \citet{kato_adaptive_2025}.};
    \draw[arr] (adoptpoint) -- (adoptlocal);
    \draw[arr] (adoptpoint) -- (adoptadaptive);

    \node[box=4.1cm] (partial_data) at (22.5,4.6) {Can I learn \textit{where} to collect new data to achieve better internal and external validity?\newline \citet{opocher_geom_2026}};
    \draw[arr] (adoptpartial) -- (partial_data);
    \node[box=4.2cm] (multiple) at (21,7.0) {What if I have to decide how to marginally change existing policies? \citet{moon2026}.};
    \draw[arr] (adoptpoint) -- (multiple);
    \node[box=4.2cm] (adoptpartialleaf) at (26.8,2.0) {What if my $(\omega, \Theta_\omega, \mathcal{M})$ is only \textit{partially internally} valid?\newline \citet{stoye_covariates_2012}; \citet{yata_2025}; \citet{christensen_payoffs_2025}; \citet{Olea_2026}.};
    \draw[arr] (adoptpartial) -- (adoptpartialleaf);
    \node[box=4.1cm] (partial_ext) at (27.8,5.0) {What if my $(\omega, \Theta_\omega, \mathcal{M})$ is not \textit{externally} valid?\newline \citet{adjaho_external_2022}; \citet{kido_distributionally_2022}.};
    \draw[arr] (adoptpartial) -- (partial_ext);
\end{tikzpicture}
}

\begin{minipage}{0.92\linewidth}
\scriptsize
\emph{Note.} This decision tree provides theory-driven answers to research design questions that applied researchers may have when interested in providing (unconstrained) policy recommendations.
\end{minipage}
\end{figure}
\end{landscape}

\begin{landscape}
\begin{figure}[p]
\centering
\caption{Designing Research for Constrained Policy Choice}
\label{fig:constrained_tree}
\resizebox{0.98\linewidth}{!}{%
\begin{tikzpicture}[
    x=0.95cm,
    y=1cm,
    arr/.style={-Latex, thick},
    root/.style={draw, rounded corners, very thick, align=center, fill=blue!10, text width=5.0cm, minimum height=1.1cm, inner sep=5pt, font=\normalsize},
    branch/.style={draw, rounded corners, thick, align=center, fill=blue!6, text width=3.6cm, minimum height=1cm, inner sep=5pt, font=\small},
    box/.style={draw, rounded corners, thick, align=left, fill=white, text width=#1, inner sep=4pt, font=\scriptsize},
    box/.default=3.0cm
]
    \node[root] (root) at (12.8,14.8) {Choice for $(\omega, \Theta_\omega)$};

    \node[branch] (assignment) at (6.5,13.2) {Assignment};
    \node[branch] (policy) at (12.8,12.3) {Policy Space};
    \node[branch] (welfare) at (21.1,13.2) {Welfare Objective};
    \draw[arr] (root) -- (assignment);
    \draw[arr] (root) -- (policy);
    \draw[arr] (root) -- (welfare);

    \node[box=3.2cm] (obs) at (2.4,7.2) {What if I only have access to an observational data set?\newline \citet{athey_policy_2021}.};
    \draw[arr] (assignment) -- (obs);
    \node[box=3.2cm] (exp) at (2.4,10.2) {What if I can run a randomized experiment?\newline \citet{kitagawa_who_2018}.};
    \node[box=3.2cm] (net) at (6.6,9.2) {What if I expect interventions to generate network spillovers?\newline \citet{viviano_policy_2024}.};
    \draw[arr] (assignment) -- (exp);
    \draw[arr] (assignment) -- (net);

    \node[box=4.1cm] (actions) at (14.0,3.5) {What if policies are dynamic?\newline \citet{Shosei_2025}.};
    \draw[arr] (policy) -- (actions);
    \node[box=4.5cm] (proxy) at (12.0,5) {How should I trade the quantity and quality of data I can collect? \newline \cite{opocher_2026}};
    \draw[arr] (policy) -- (proxy);
    \node[box=5.5cm] (complexity) at (5.6,4.2) {How complex should the policy class be?\newline \citet{mbakop_model_2021}; \citet{Ponomarev2026}.};
    \draw[arr] (policy) -- (complexity);
    \node[box=4.0cm] (specialized) at (10.8,7.2) {What if interventions are not binary?\newline \citet{fang_2025}; \citet{ai_2026}.};
    \draw[arr] (policy) -- (specialized);
    \node[box=4.1cm] (laissez) at (16.5,8.5) {What if I want to let people self-select?\newline \citet{kitagawa_choosing_who_choses}.};
    \draw[arr] (policy) -- (laissez);
    \node[box=4.5cm] (match) at (16.5,10.5) {What if I consider matching policies?\newline \citet{hazard_kitagawa_2025}.};
    \draw[arr] (policy) -- (match);

    \node[box=5.5cm] (ineq) at (18.2,5.7) {What if I care about inequality?\newline \citet{kitagawa_equality-minded_2021}; \citet{fan_qi_xu_2025}; \citet{terschuur_2025}.};
    \draw[arr] (welfare) -- (ineq);
    \node[box=4.0cm] (randomc) at (23.8,6.2) {What if feasibility constraints are unknown?\newline \citet{sun_2026}.};
    \draw[arr] (welfare) -- (randomc);
    \node[box=4.0cm] (fair) at (21.5,9.2) {What if I care about fairness across protected groups?\newline \citet{viviano_fair_2024}.};
    \draw[arr] (welfare) -- (fair);
\end{tikzpicture}
}

\begin{minipage}{0.92\linewidth}
\scriptsize
\emph{Note.} This decision tree provides theory-driven answers to research design questions that applied researchers may have when interested in providing (constrained) policy recommendations.
\end{minipage}
\end{figure}
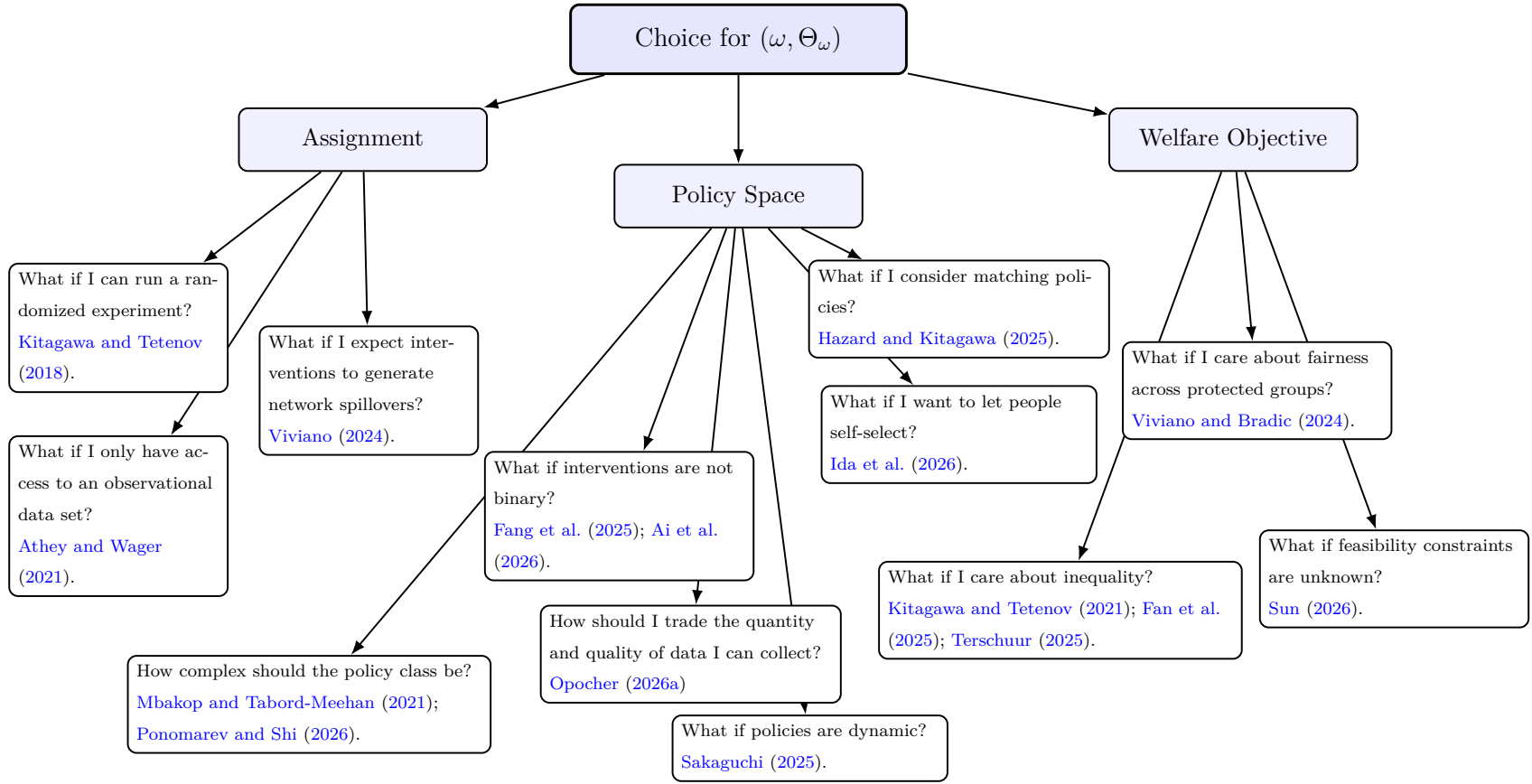
\end{landscape}

\subsection{\textit{After} Collecting Data}\label{sec:after_collecting}
In this section, we focus on the stage where data collection and the main analysis have already been completed.
Therefore, the main focus here switches from making research design choices to communicating a policy recommendation to the policymaker.
In particular, the main objective of this section is to provide applied researchers with \textit{two figures} and \textit{a table} they can plug in the `policy implications' section of their applied paper to provide evidence on the expected performance of their recommendation.

Let's suppose a researcher has collected data from the randomized controlled trial in \citet{hussam_2022}, which was funded with the objective of producing recommendations about whether to distribute the intervention through full adoption, no adoption, or targeting based on prespecified covariates with a rectangular rule. 
For simplicity, let us abstract from the design challenges highlighted in the previous section and focus on the simplest setting, which is the one the authors considered.

The rankings entrepreneurs give one another for future profits, henceforth community rankings, play a central role in the paper.
\citet[][pp. 880]{hussam_2022} write: 
`[...] lending institutions or other organizations aiming to target capital to entrepreneurs with productive opportunities would have good reason to leverage community information'.
\citet[][pp.~879--880]{hussam_2022} motivate this conclusion with the targeting results in Table 2. 
In the discussion immediately below Table 2, they note that bottom-tercile entrepreneurs have returns statistically indistinguishable from zero, middle-tercile effects are not statistically significant, and the strongest intervention effects are concentrated in the top tercile. 
They also emphasize that these estimates are stable after controlling for a rich set of observables. 
Consistent with this, Table 4 shows that adding community rankings to observables strengthens prediction.
Although these findings provide valid evidence for discovery objectives alone, they do not provide direct evidence in support of the normative objective: 
should the PM roll out the cash transfer under full adoption? 
Should the PM shut down the policy and assign it to nobody? 
Or is targeting based on the community rankings the best option?

To answer these questions, I introduce (the beta version of) a new \texttt{R} package called \texttt{policytargetr} that takes the outcome variable, the intervention dummy, the propensity score, and two targeting covariates as inputs and produces (i) a plot of the estimated rectangular frontier, (ii) a plot of the welfare produced by each recommendation, and (iii) a companion table.

Listing \ref{lst:policytargetr_workflow} reports the workflow used in the context of \citet{hussam_2022}.
The first step is to clean the application data so that it matches the package's expected format: one row per unit and four required columns called \texttt{outcome}, \texttt{intervention status}, \texttt{x1}, and \texttt{x2}. In this application, \texttt{outcome} is follow-up profits, \texttt{treatment} is the treatment-assignment dummy, and \texttt{x1} and \texttt{x2} are, respectively, age and community-rank percentile, the two targeting variables.

\begin{lstlisting}[style=policytargetr,caption={How to use \texttt{policytargetr}},label={lst:policytargetr_workflow}]
library(policytargetr)

analysis_data <- data.frame(
  outcome = raw_data$profits,
  treatment = raw_data$treat,
  x1 = raw_data$age,                    # age
  x2 = raw_data$Quintile_Rank_NS_Pct    # community rank percentile
)

fit <- analyze_rectangular_policy(
  data = analysis_data,
  propensity = 1 / 3,
  train_share = 0.6,
  split_seed = 2020,
  cost_fun = normalized_grant_cost
)

save_policy_outputs(
  fit,
  output_dir = "output_hussam_age_rank",
  prefix = "hussam_age_rank",
  frontier_x_label = "Age",
  frontier_y_label = "Rank"
)
\end{lstlisting}

\begin{figure}[H]
\centering
\caption{Out-of-Sample Empirical Welfare of each Recommendation}
\includegraphics[width=0.95\linewidth]{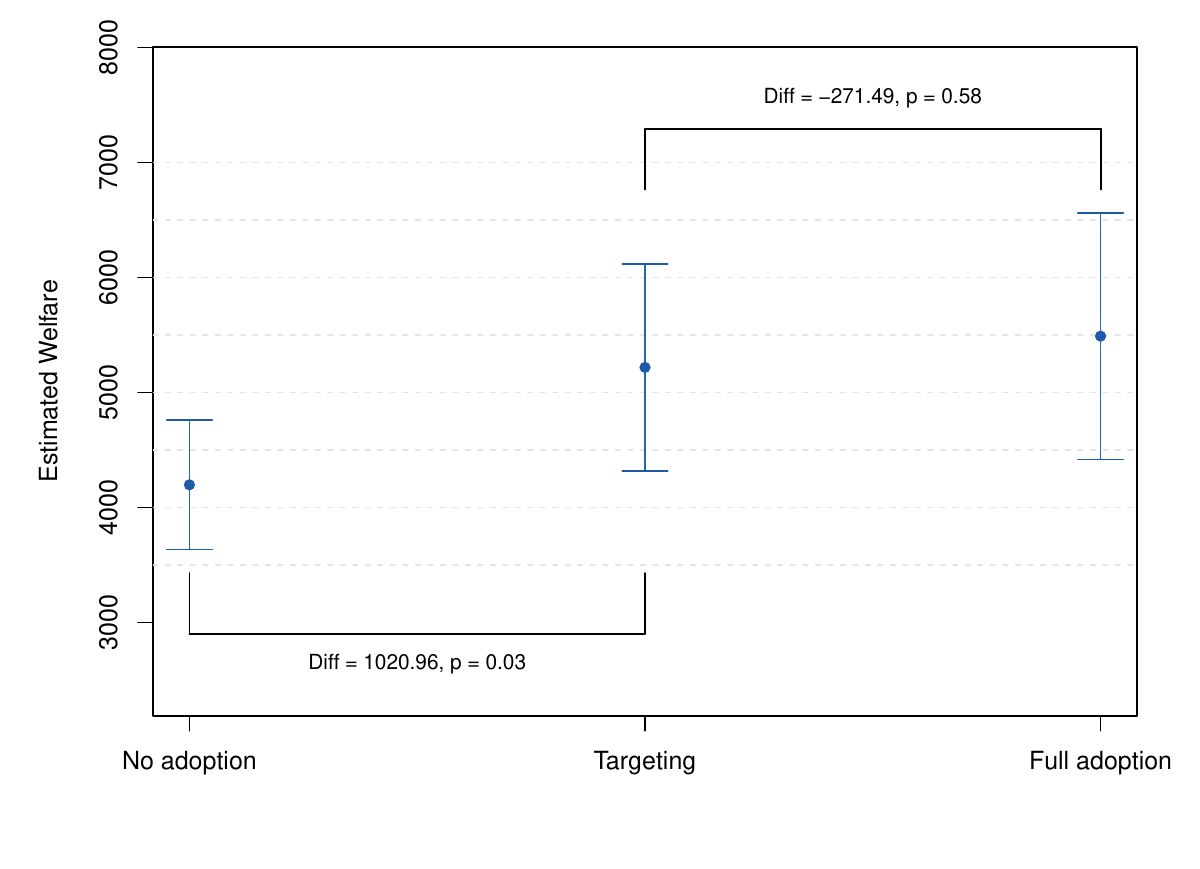}
\label{fig:hussam_policy_values}
\begin{minipage}{1\linewidth}
\footnotesize
\textbf{\textit{Notes}}. The figure reports out-of-sample empirical welfare with 95\% confidence intervals for no adoption, targeting, and full adoption. 
The brackets summarize the pairwise comparisons between targeting, full adoption, and no adoption.
\end{minipage}
\end{figure}

\begin{figure}[H]
\centering
\caption{Estimated Targeting Frontier}
\includegraphics[width=0.95\linewidth]{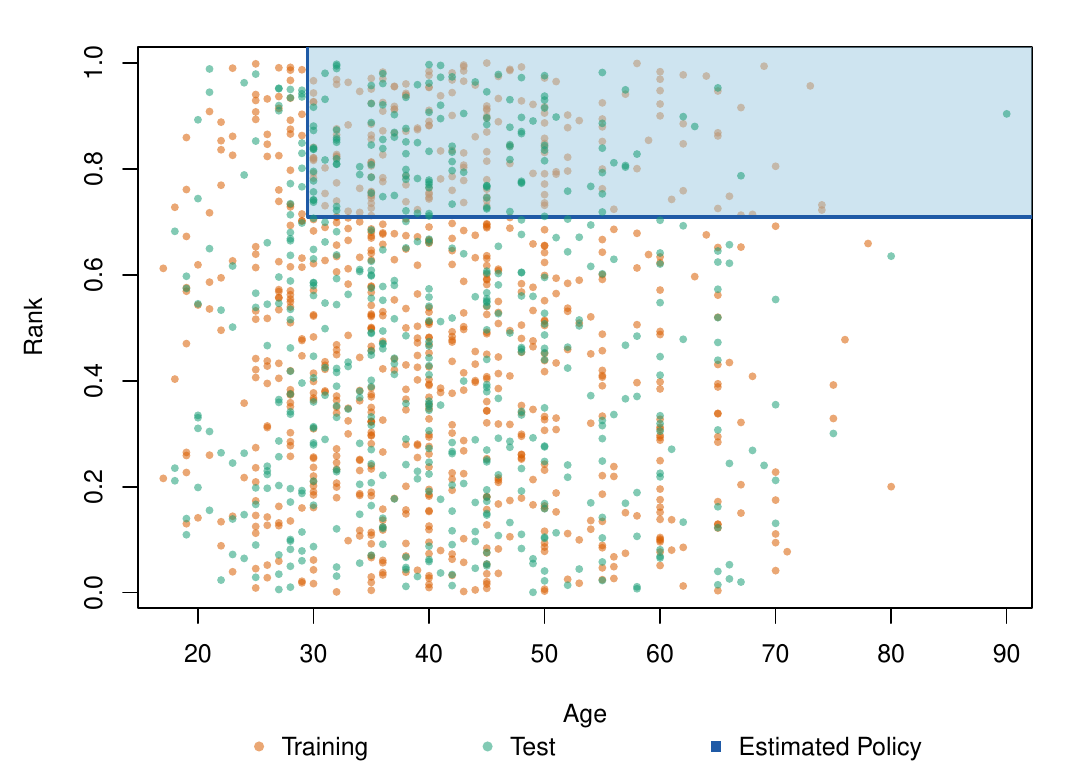}
\label{fig:hussam_frontier}
\vspace{0.5cm}
\begin{minipage}{1\linewidth}
\footnotesize
\textbf{\textit{Notes}}. The figure overlays the estimated rectangular policy frontier on the covariate scatter, with training and test observations shown in different colors. 
The axes correspond to age and community rank percentile. The shaded area indicates assignment to receive the cash transfer.
\end{minipage}
\end{figure}

In a nutshell, the package uses \texttt{split\_seed} and \texttt{train\_share} to split the original data set into training and test sets.
One can also specify a policy cost function (the $\sigma$ function in Definition \ref{def:policy_choice_problem}) with the argument \texttt{cost\_fun}.
For this specific application we know each grant costs 6,000 INR. 
To ease interpretation, we normalize the cost of treating the whole test set to 1.
On the training sample, \texttt{analyze\_rectangular\_policy()} takes \texttt{propensity} and the formatted data object as inputs to learn the EWM \citep{kitagawa_who_2018} rectangular targeting rule. 
It then computes the empirical welfare under no adoption, full adoption, and targeting on the test sample, returning the fitted object \texttt{fit}.
Because the training and test samples are independent, we can perform valid holdout inference and provide standard errors and confidence intervals over the standard Normal approximation together with out-of-sample empirical welfare.
Note that the literature proposed more sophisticated ways of performing inference on welfare \citep[e.g.][]{ponomarev2025lowerconfidencebandoptimal}, which will be incorporated in the next versions of the package.
With \texttt{save\_policy\_outputs}, the practitioner saves Table \ref{tab:synthetic_honest_split} and Figures \ref{fig:hussam_policy_values} and \ref{fig:hussam_frontier} in the \texttt{output\_dir}, with prespecified labels for graphs and a prefix for file names.

Of course, one can extend this beta version, which can be considered a minimum viable product, in many ways by following the literature reviewed in this paper.
The same output could then be produced under various departures from this simple setting that follow the directions highlighted in Figures \ref{fig:unconstrained_tree} and \ref{fig:constrained_tree}.

Table \ref{tab:synthetic_honest_split} provides a summary of the results. 
Panel A reports general information: the total sample size, the experimental propensity score, and, most importantly, the estimated targeting rule, which assigns the cash grant to entrepreneurs older than 29.5 years and above the 71st percentile of the community ranking. 
Panel B then reports, for each recommendation, both empirical welfare and normalized implementation cost. 
Full adoption attains the highest point estimate, but targeting is close in welfare while costing one-fourth as much under the specified cost function. 
Panel C is the key decision panel: targeting dominates no adoption by about 1,021 INR in holdout welfare and this difference is statistically significant, whereas the difference between targeting and full adoption is small and imprecise. 
The implication is that using community information to target the grant improves meaningfully on the status quo and preserves most of the gains from universal rollout at a much lower cost.

\begin{table}[htbp]
\centering
\caption{Choosing Across Recommendations}
\label{tab:synthetic_honest_split}
\scalebox{0.9}{
\begin{tabular}{llllll}
\toprule
\multicolumn{6}{l}{\textbf{Panel A. General Information}}\\
\midrule
Sample Size & \multicolumn{5}{l}{1329}\\
Propensity Score & \multicolumn{5}{l}{0.33}\\
Estimated Policy & \multicolumn{5}{l}{$\hat d(x)=\mathbf{1}\{x_1 > 29.50,\ x_2 > 0.71\}$}\\
\addlinespace[0.5em]
\multicolumn{6}{l}{\textbf{Panel B. Policy Values}}\\
\midrule
Policy & Estimate & Cost & Std. Error & \multicolumn{2}{l}{95\% CI}\\
No adoption & 4197.49 & 0.00 & 286.51 & \multicolumn{2}{l}{[3634.65, 4760.33]} \\
Full adoption & 5489.94 & 1.00 & 545.24 & \multicolumn{2}{l}{[4418.84, 6561.04]} \\
Targeting & 5218.45 & 0.25 & 458.41 & \multicolumn{2}{l}{[4317.93, 6118.96]} \\
\addlinespace[0.5em]
\multicolumn{6}{l}{\textbf{Panel C. Policy Comparisons}}\\
\midrule
Comparison & Gains & Cost Diff. & Std. Error & 95\% CI & p-value\\
Targeting - No adoption & 1020.96 & 0.25 & 473.43 & [90.93, 1950.98] & 0.03 \\
Targeting - Full adoption & -271.49 & -0.75 & 493.04 & [-1240.04, 697.05] & 0.58 \\
Full adoption - No adoption & 1292.45 & 1.00 & 682.77 & [-48.81, 2633.72] & 0.06 \\
\bottomrule
\end{tabular}}
\begin{minipage}{1\linewidth}
\vspace{0.5em}
\footnotesize
\textbf{\textit{Notes.}} Panel B reports test set empirical welfare under adoption, full adoption, and targeting. 
Panel C reports paired test set emprical welfare differences across the three pairwise policy comparisons, together with the corresponding cost differences. 
The Panel C p-values test the null hypothesis that each hold-out welfare difference is zero.
\end{minipage}
\end{table}

Figure \ref{fig:hussam_policy_values} visualizes the comparison in Panels B and C of Table \ref{tab:synthetic_honest_split}. 
The dots report test-set empirical welfare for the three candidate recommendations, the vertical bars report 95\% confidence intervals, and the brackets add the pairwise welfare differences together with their \textit{p}-values. 
First, the targeting rule lies clearly above no adoption, and the lower bracket makes it transparent that this gain is statistically significant. 
Second, targeting and full adoption are visually very close, with overlapping confidence intervals and a small, insignificant bracketed difference. 
Thus, the figure conveys the main practical message of the application immediately: targeting appears to capture most of the welfare gains from expansion while avoiding the cost of treating everyone.

Figure \ref{fig:hussam_frontier} visualizes \textit{who} should receive the transfer, as also reported in Table \ref{tab:synthetic_honest_split}, Panel A. 
The blue vertical and horizontal cutoffs trace the estimated rectangular policy frontier, while the shaded upper-right region identifies the entrepreneurs who receive the grant under the recommended policy. 

\begin{remark}
    Note that in the current version, this package has limited application scope. 
    Here I highlight some information that the results do \textbf{not} provide.
    First, the output is not informative about what would happen if the PM were to apply the policy recommendation to a population that is different from the one in the study, say African entrepreneurs.
    Second, the output is not informative about any intervention that differs from the one considered in the study. 
    Therefore, the PM should replicate the authors' implementation step by step. 
    Third, any violation of the experimental design could undermine the validity of the normative exercise. 
    We can use Figure \ref{fig:constrained_tree} to navigate the departures from the simple design considered here.
    The same applies to the policy space and the welfare objective.
\end{remark}

\section{Conclusions}
Economists' role is recognized worldwide as providing empirical evidence to guide real-world policy choices. 
Public institutions, research organizations, and the profession itself acknowledge this role. 
As a result, much empirical work in economics is motivated, and often funded, by broader normative objectives. 
The success in achieving such objectives is tied to economists' \textit{research design} choices.

In this review, I first surveyed the theoretical literature initiated by \citet{manski_statistical_2004} that studies this problem through the lens of statistical decision theory \citep{wald1950statistical, Savage01031951}. 
I proposed a common framework that nests choices over data collection, design margins, estimators, and decision rules, and used it to organize the literature into unconstrained and constrained policy-choice problems. 
This made it possible to clarify under a common framework what different papers hold fixed, what object they optimize, what type of regret guarantee they deliver, and how recent work extends the earlier contributions.

Then, I turned to the applied side and provided two tools that researchers can use when the goal is to produce and communicate policy recommendations. 
First, I introduced a practitioner-oriented guide for the \textit{ex-ante} stage, meant to help researchers write a normative pre-analysis plan by mapping concrete design concerns into the relevant branch of the theory. 
Second, I introduced a simple \textit{ex-post} workflow, operationalized through the \texttt{policytargetr} package, to help communicate to policymakers not only \textit{what} the recommended policy is, but also how it compares with alternatives in terms of welfare and implementation costs. 

This review ultimately aims to organize recent theoretical results under a common framework and bring them closer to mainstream empirical practice.

\bibliography{bib}
\clearpage

\appendix
\setcounter{table}{0}
\renewcommand{\thetable}{A\arabic{table}}
\setcounter{figure}{0}
\renewcommand{\thefigure}{A\arabic{figure}}

\section{Additional Tables and Figures}
\begin{table}[H]
\centering
\caption{Mapping the general framework into applied research language}
\label{app:tab:notation_applied}
\begin{tabular}{p{1.4cm} p{4.2cm} p{8.0cm}}
\toprule
Notation & Jargon & Definition and Examples \\
\midrule
$\omega \in \Omega$ & Data-collection process & What data are collected and how: for example, a randomized experiment, an observational sample, a survey, administrative records, fixed or adaptive assignment. \\
$\theta \in \Theta_\omega$ & Collection margins & The margins of choice within the collection process: sample size, propensity scores, subgroup-specific sampling shares, follow-up length, types of measurements, variable selection. \\
$m \in \mathcal{M}$ & Estimator & The object used to attach a value to each policy using data: for example, a difference in means, a doubly robust score, a DiD estimator, empirical welfare. \\
$\delta \in \Delta$ & Decision rule & The rule that turns estimated values into a recommendation: for example, treat if $\hat\tau>0$, target groups with positive estimated gains, rank policies under a budget, or choose the policy with the highest estimated welfare. \\
\bottomrule
\end{tabular}
\end{table}

\section{Counting Recommendations in the Applied Literature}\label{app:intro_counts_methodology}
To construct the descriptive statistics reported in the introduction, I assembled the universe of papers published between 2015 and 2025 in \textit{American Economic Review}, \textit{AER: Insights}, \textit{AEJ: Applied Economics}, \textit{AEJ: Economic Policy}, and \textit{AER: Papers \& Proceedings}, and coded each paper from its full text using the following procedure. 
First, each paper was classified according to whether its main contribution was empirical. 
Second, among empirical papers, I classified whether the paper contained at least one policy recommendation. 
Third, whenever a paper contained more than one distinct recommendation, each recommendation was recorded separately, so that paper counts and recommendation counts are allowed to differ. 

The classification was performed using OpenAI's API and the \texttt{gpt-5.2} model. 
The exact instruction prompt is reproduced below. 
Notice that it also gathered other information that is not used in the counts reported in the introduction.

\begin{lstlisting}[basicstyle=\ttfamily\footnotesize,breaklines=true]
You are coding one economics paper for a study of empirical practice in policy recommendations.
Read the full paper and return exactly one JSON object that matches the supplied schema.
Each row should be a paper-recommendation pair.
Core rules:
1. Be conservative. Do not infer a recommendation unless the paper supports it.
2. Separate paper-level coding from recommendation-level coding.
3. If a paper has more than one distinct recommendation, code each separately.
4. Prefer an empty list over guessing.
5. Use brief evidence quotes only, and cite page markers like [[PAGE 12]] when available.

Stage 1. Main contribution empirical? (JSON main_contribution_empirical)
- Code "Yes" if the paper's main novelty is empirical evidence from data on a real-world, policy-relevant question.
- Code "No" if the paper is mainly theoretical, methodological, a review, or otherwise not mainly empirical.
Also extract:
- the journal name
- the list of authors
If either is not clearly visible in the text, use "unclear".
IF "NO", GO TO NEXT PAPER.
Stage 2. Recommendation present? (JSON policy_recommendation)
A recommendation is any claim about:
- who should receive treatment (who)
- whether an intervention or policy should be adopted, expanded, continued, or stopped (whether)
- how much to invest, scale, or allocate (how_much)
- "Explicit": the paper clearly recommends an action.
- "Implicit": the paper does not directly recommend an action, but clearly suggests one.
- "No recommendation": the paper reports findings without moving to an action claim.
IF "NO RECOMMENDATION", GO TO NEXT PAPER.

Stage 3. For EACH distinct recommendation, generate a row.
For EACH coded recommendation, report:
- the section title where the recommendation appears: "title", "abstract", "introduction", "main_results" (section where main results are displayed), "policy" (if there is a separate section for policy recommendations), "conclusions".
- the explicit sentence that contains the recommendation; if there is no single sentence, quote the shortest passage that most clearly states it.

Moreover, report: (Direct correspondence with JSON):
- D: "who", "whether", "how_much", "other". 
- m: "Regression Discontinuity Design", "Difference in Differences", "Instrumental Variables", "Linear Regression", "Inverse Probability Weighting", "Difference in Means", "Structural Estimation", "Machine Learning Prediction".
- delta: "significance_rule" (point estimate + significant/non-significant), "structural_ranking_rule" (highest point estimate across counterfactuals), "unclear", "other".
- omega: "survey experiment", "lab experiment", "field experiment", "panel data", "cross-sectional data", "time series data", "other".

Evidence rules:
- Use only information contained in the paper.
- Keep quotes short.
- The recommendation sentence should be copied verbatim from the paper when possible.
- If the text is incomplete, corrupted, or too ambiguous to code reliably, set needs_manual_review to true and explain why.
- Guess as little as possible 

Output rules:
- Return JSON only.
- Do not wrap the JSON in markdown.
- Do not add commentary before or after the JSON.
\end{lstlisting}

\end{document}